\documentclass[draftclsnofoot,onecolumn,12pt]{IEEEtran}

\usepackage[nospace,noadjust]{cite}

\usepackage{amssymb,amsmath}
\usepackage{graphics}
\usepackage{float}
\usepackage{graphicx}
\usepackage{amsmath}
\usepackage{cite}
\usepackage{color}
\usepackage{dsfont}
\usepackage{bm}
\usepackage{upgreek}
\usepackage{arydshln}
\usepackage{enumerate}
\usepackage{amsthm}
\usepackage{graphicx}
\usepackage{threeparttable}
\usepackage{caption}
\usepackage{multirow}
\usepackage{color}
\usepackage{xfrac}
\usepackage{array}
\usepackage{nomencl}
\usepackage{hyphenat}

\newcommand{\be}{\begin{equation}}
\newcommand{\ee}{\end{equation}}

\newcommand{\parder}[2]{ \frac{\partial #1}{\partial #2} }
\newcommand{\norm}[1]{ || #1 ||}
\newcommand{\mb}[1]{\mathbf{#1}}
\newcommand{\bs}[1]{\boldsymbol{#1}}
\newcommand{\lbr}{\left\lbrace}
\newcommand{\rbr}{\right\rbrace}
\newcommand{\virg}[1]{\textquotedblleft#1\textquotedblright}

\newcommand{\vsigmat}{\mathrm{vecs}(\bs{\Sigma}_0)}
\newcommand{\vcsigma}{\mathrm{vec}(\bs{\Sigma})}
\newcommand{\vcsigmatinv}{\mathrm{vec}(\bs{\Sigma}_0^{-1})}

\newcommand{\kronsigmatinvT}{\bs{\Sigma}_0^{-T}\otimes\bs{\Sigma}_0^{-1}}

\newcommand{\kronsigmatinvmT}{\bs{\Sigma}_0^{-T/2}\otimes\bs{\Sigma}_0^{-1/2}}

\newcommand{\vecl}[1]{\mathrm{vec}_l(#1)}

\newcommand{\cvec}[1]{ \mathrm{vec}\left(  #1 \right) }
\newcommand{\tonde}[1]{\left( #1 \right)  }
\newcommand{\quadre}[1]{\left[  #1 \right]  }
\newcommand{\graffe}[1]{\left\lbrace   #1 \right\rbrace   }

\newtheorem{definition}{Definition}[section]

\begin{document}
	
\title{Semiparametric CRB and Slepian-Bangs formulas for Complex Elliptically Symmetric Distributions}

\author{Stefano~Fortunati,~\IEEEmembership{Member,~IEEE,}
	Fulvio~Gini,~\IEEEmembership{Fellow,~IEEE,}
	Maria~S.~Greco,~\IEEEmembership{Fellow,~IEEE,}
	Abdelhak~M.~Zoubir,~\IEEEmembership{Fellow,~IEEE}
	and~Muralidhar~Rangaswamy,~\IEEEmembership{Fellow,~IEEE} % <-this % stops a space
	\thanks{S. Fortunati is with Universit\`a di Pisa, Dipartimento di Ingegneria dell'Informazione, Pisa, Italy and with Technische Universit\"at Darmstadt, Signal Processing Group, Darmstadt, Germany 
		(e-mail: stefano.fortunati@iet.unipi.it).}% <-this % stops a space
	\thanks{F. Gini and M. S. Greco are with Universit\`a di Pisa, Dipartimento di Ingegneria dell'Informazione, Pisa, Italy (e-mail: f.gini,m.greco@iet.unipi.it).}% <-this % stops a space
	\thanks{A. M. Zoubir is with Technische Universit\"at Darmstadt, Signal Processing Group, Darmstadt, Germany (e-mail: zoubir@spg.tu-darmstadt.de).}
	\thanks{M. Rangaswamy is with U.S. AFRL, Sensors Directorate, Wright-Patterson AFB, OH, USA (e-mail: muralidhar.rangaswamy@us.af.mil).}}

\maketitle

\begin{abstract}
The main aim of this paper is to extend the semiparametric inference methodology, recently investigated for Real Elliptically Symmetric (RES) distributions, to Complex Elliptically Symmetric (CES) distributions. The generalization to the complex field is of fundamental importance in all practical applications that exploit the complex representation of the acquired data. Moreover, the CES distributions has been widely recognized as a valuable and general model to statistically describe the non-Gaussian behaviour of datasets originated from a wide variety of physical measurement processes. The paper is divided in two parts. In the first part, a closed form expression of the constrained Semiparametric Cram\'{e}r-Rao Bound (CSCRB) for the joint estimation of complex mean vector and complex scatter matrix of a set of CES-distributed random vectors is obtained by exploiting the so-called \textit{Wirtinger} or $\mathbb{C}\mathbb{R}$-\textit{calculus}. The second part deals with the derivation of the semiparametric version of the Slepian-Bangs formula in the context of the CES model. Specifically, the proposed Semiparametric Slepian-Bangs (SSB) formula provides us with a useful and ready-to-use expression of the Semiparametric Fisher Information Matrix (SFIM) for the estimation of a parameter vector parametrizing the complex mean and the complex scatter matrix of a CES-distributed vector in the presence of unknown, nuisance, density generator. Furthermore, we show how to exploit the derived SSB formula to obtain the semiparametric counterpart of the Stochastic CRB for Direction of Arrival (DOA) estimation under a random signal model assumption. Simulation results are also provided to clarify the theoretical findings and to demonstrate their usefulness in common array processing applications.   
\end{abstract}


\section{Introduction}
\label{complex_data}

Statistical analysis of \textit{complex} data is a well-established field in Signal Processing (see \cite{Brad, Pici, Pici96, Olede, Sharf, Erik, Adali, draskovic} just to cite a few). The use of complex representation for the acquired data can simplify the modeling and the inference tasks in many applications such as acoustics, optics, seismology, communications and radar/sonar Signal Processing. This fact, together with the need to model the non-Gaussian, heavy-tailed statistical behavior of the disturbance, led to the introduction of the wide family of Complex Elliptical Symmetric (CES) distributions (\cite{CES_stat}, \cite[Ch. 3]{RichPhD}, \cite{zozor}, \cite{Esa}, \cite{Grec3} and \cite[Ch. 4]{book_zoubir}). Briefly, if an $N$-dimensional complex random vector $\mb{z} \in \mathbb{C}^N$ is CES-distributed, say $\mb{z} \sim CES_N(\bs{\mu},\bs{\Sigma}, h)$, then its probability density function (pdf) is fully specified by the complex mean vector $\bs{\mu} \in \mathbb{C}^N$, the complex scatter matrix $\bs{\Sigma} \in \mathbb{C}^{N \times N}$ and the density generator $h \in \mathcal{G}$, where $\mathcal{G}$ is a suitable set of functions. CES distributions are the complex extension of Real Elliptically Symmetric distributions \cite{CAMBANIS1981, RES_Fang} from which they inherit most of their properties.

Our recent papers \cite{For_EUSIPCO, For_SCRB} focuses on the particular semiparametric\footnote{The reader that is not familiar with the semiparametric theory may have a look at the books \cite{BKRW} and \cite{Tsiatis} or to the wide statistical literature available on this topic and partially collected in the reference lists of \cite{For_EUSIPCO,For_SCRB}.} structure of the RES distributions. As noted in \cite{Bickel_paper} and \cite[Sec. 4.2 and 7.2]{BKRW}, the RES distributions can be considered as a semiparametric group model whose parametric part is given by the mean vector and by the scatter matrix to be jointly estimated, while the non-parametric \textit{nuisance} part is given by the density generator. Moreover, in \cite{For_SCRB}, a closed form expression for the Semiparametric Cram\'{e}r-Rao Bound (SCRB) on the joint estimation of the parametric part of the RES model has also been derived. It is worth noticing that the SCRB for the estimation of the mean vector and of the scatter matrix has been already derived in \cite{Hallin_P_Annals, Hallin_Annals_Stat_2, Hallin_P_2006, PAINDAVEINE} by using a more general, but more abstract, procedure based on the LeCam's theory \cite{LeCam}.

The aim of this paper is to generalize and extend the results on the SCRB, already derived in the context of RES distributions in \cite{For_SCRB}, to CES distributions. Firstly, we will provide a closed form expression for the SCRB on the Mean Square Error (MSE) of the joint estimation of the \textit{complex} mean vector $\bs{\mu}$ and \textit{complex} scatter matrix $\bs{\Sigma}$ of a set of CES distributed random vectors. This generalization relies on the Wirtinger or $\mathbb{C}\mathbb{R}$-calculus (\cite{Bos_Grad, Remmert, Sharf, Erik, Adali, Li, Kreutz}) and on its application on the derivation of lower bounds (\cite{Bos, Rao, OllCRB, Menni, Rich, Complex_MCRB_For}). Then, the second part of the paper is dedicated to the derivation of a semiparametric version of the the celebrated Slepian-Bangs (SB) formula and the related Semiparametric Stochastic CRB (SSCRB) for Direction of Arrival (DOA) estimation problems.  

Introduced by Slepian and Bangs in \cite{Slepian} and \cite{Bangs}, the SB formula has been extensively used for many years in array processing. The \virg{classic} SB formula is a compact expression of the Fisher Information Matrix (FIM) for parameter estimation under a Gaussian data model \cite[Appendix 3C]{kay1993fundamentalsI}. Specifically, let $\bs{\theta} \in \Theta \subset \mathbb{R}^d$ be a $d$-dimensional, deterministic parameter vector and let $\mathbb{C}^{N} \ni \mb{z} \sim CN(\bs{\mu}(\bs{\theta}),\bs{\Sigma}(\bs{\theta}))$ be a possibly complex, Gaussian-distributed, random vector (also called \textit{snapshot}), representing the available observation. Then the SB formula provides us with a closed-form expression of the FIM for the estimation of $\bs{\theta} \in \Theta$. 

Due to its central role in many practical applications, including DOA estimation, the SB formula has been the subject of active research. In particular, it has been generalized to non-Gaussian and \textit{mismatched} estimation frameworks \cite{SPM}. Specifically, in \cite{Bess}, Besson and Abramovich proposed a generalization of the classical, SB formula to CES-distributed data. Richmond and Horowitz in \cite{Rich} showed an extension of the classical, Gaussian-based, SB formula to estimation problems under model misspecification. The natural follow-on \cite{Bess} and \cite{Rich} has been proposed in \cite{miss_sb}, where SB-type formulas, that encompass those previously obtained in \cite{Bess} and \cite{Rich} as special cases, have been derived for parameter estimation problems involving CES-distributed data under model misspecification. In this paper, we take a step forward to the generalization of the SB formula for semiparametric estimation in the CES framework. Concretely, we propose a Semiparametric SB (SSB) formula that provides a compact expression of the Semiparametric FIM (SFIM) for the estimation of $\bs{\theta} \in \Theta$ in CES-distributed data when the density generator is unknown. More specifically, let $\mathbb{C}^N \ni \mb{z} \sim CES_N(\bs{\mu}(\bs{\theta}),\bs{\Sigma}(\bs{\theta}), h)$ be a CES-distributed random vector parameterized by $\bs{\theta} \in \Theta \subset \mathbb{R}^d$, then the SCRB related to the proposed SSB formula provides a lower bound on the Mean Square Error (MSE) of \textit{any} estimator of $\bs{\theta}$ in the presence of an \textit{unknown}, nuisance density generator $h \in \mathcal{G}$. It is worth pointing out that, we assume here the unknown parameter vector $\bs{\theta} \in \Theta$ to be real-valued since in most of the practical application of the SSB formula $\bs{\theta}$ collects real parameters (e.g. the DOAs of a certain number of sources in array processing). This assumption, however, does not represent a limitation, since we can always maps a complex vector in a real one simply by stacking its real and the imaginary parts. Moreover, Wirtinger calculus may be exploited to obtain the proposed SSB formula directly in the complex field. 
We conclude the paper with an example of application of the derived SSB formula. In particular, we provide a closed form expression of the so-called \virg{Stochastic} CRB for the DOA estimation in the presence of a random signal model \cite{Stoica_CRB,Stoica_CRB_2,Ottersten,Weiss,Renaux}.  

\textit{Notation}: Throughout this paper, italics indicates scalar quantities ($a$), lower case and upper case boldface indicate column vectors ($\mathbf{a}$) and matrices ($\mathbf{A}$), respectively. Note that the word \virg{vector} indicates both Euclidean vectors and vector-valued functions. For the sake of clarity, we indicate sometimes a vector-valued function as $\mb{a}\equiv\mb{a}(\mb{z})$. The asterisk $*$ indicates complex conjugation. The superscripts $T$ and $H$ indicate the transpose and the Hermitian operators respectively, then ${{\mathbf{A}}^H} = {({{\mathbf{A}}^ * })^T}$. Moreover, $\mb{A}^{-T} \triangleq (\mb{A}^{-1})^T = (\mb{A}^T)^{-1}$, $\mb{A}^{-*} \triangleq (\mb{A}^{-1})^* = (\mb{A}^*)^{-1}$ and $\mb{A}^{-H} \triangleq (\mb{A}^{-1})^H = (\mb{A}^H)^{-1}$. Each entry of a matrix $\mb{A}$ is indicated as $a_{i,j}\triangleq [\mb{A}]_{i,j}$. Let $\mb{A}(\bs{\theta})$ be a matrix (or possibly a vector or even a scalar) function of the \textit{real} vector $\bs{\theta} \in  \mathbb{R}^d$, then $\mb{A}_0 \triangleq \mb{A}(\bs{\theta}_0)$ while $\mb{A}_i^{0}\triangleq {\frac{\partial \mb{A}(\bs{\theta})}{\partial \theta_i}}|_{\bs{\theta}=\bs{\theta}_0}$ and $\mb{A}_{ij}^{0} \triangleq \frac{\partial^2 \mb{A}(\bs{\theta})}{{\partial \theta_i}{\partial \theta_j}}|_{\bs{\theta}=\bs{\theta}_0}$, where $\bs{\theta}_0$ is a particular (or \textit{true}) value of $\bs{\theta}$. $\mb{I}_N$ defines the $N \times N$ identity matrix. According to the notation introduced in \cite{For_EUSIPCO} and \cite{For_SCRB}, we indicate the \textit{true} pdf as $p_0(\mb{z})\triangleq p_Z(\mb{z}|\bs{\theta}_0, h_0)$, where $h_0$ indicates the true nuisance function. Moreover, $E_0\{\cdot\}$ indicates the expectation operator with respect to (w.r.t.) the true pdf $p_0(\mb{z})$. Finally, for random variables or vectors, $=_d$ stands for "has the same distribution as".

\section{A brief recap on CES distributions}
\label{extension_CES}
This section provides a brief overview of CES distributions with a specific focus on the properties that will play a crucial role in the derivation of the complex version of the SCRB and the SSB formula. 
\begin{definition}
	\label{def_CES}
	(\cite{CES_stat}, \cite{RichPhD}, \cite{Esa} and \cite[Ch. 4]{book_zoubir}) Let $\mb{z} \triangleq \mb{x}_R + j\mb{x}_I \in \mathbb{C}^N$ be a complex random vector and let $\mb{x}_R \in \mathbb{R}^N$ and $\mb{x}_I \in \mathbb{R}^N$ be two real random vectors that represent the real and the imaginary part of $\mb{z}$, respectively. Then $\mb{z}$ is said to be CES-distributed with mean vector $\bs{\mu}$ and scatter matrix $\bs{\Sigma}$ such that (s.t.):
	\be
	\label{par_CES}
	\bs{\mu} = \bs{\mu}_R + j\bs{\mu}_I \in \mathbb{C}^N \quad \bs{\Sigma} = \mb{C}_1 +j\mb{C}_2 \in \mathbb{C}^{N \times N},
	\ee
	if and only if the real random vector $\tilde{\mb{x}} \triangleq (\mb{x}_R^T,\mb{x}_I^T)^T \in \mathbb{R}^{2N}$ is RES-distributed with mean vector $\tilde{\bs{\mu}} = (\bs{\mu}_R^T, \bs{\mu}_I^T)^T$ and scatter matrix $\tilde{\bs{\Sigma}}$ that satisfies the following structure
	\be
	\label{scatter_real}
	\tilde{\bs{\Sigma}}=\frac{1}{2}\left( \begin{array}{cc}
		\mb{C}_1 & -\mb{C}_2\\
		\mb{C}_2 & \mb{C}_1
	\end{array}\right).
	\ee
\end{definition}
We note that, as a consequence of Definition \ref{def_CES}, any CES-distributed random vector $\mb{z}$ satisfies the circularity property, i.e. $(\mb{z}-\bs{\mu}) =_d e^{j\vartheta}(\mb{z}-\bs{\mu}), \; \forall \vartheta \in \mathbb{R}$.
Moreover, under the \textit{absolutely continuous} case, i.e. when the scatter matrix has full rank, the pdf of the CES-distributed vector $\mb{z}$ can be directly obtained from the one of the RES-distributed vector $\tilde{\mb{x}}\sim RES_{2N}(\tilde{\mb{x}};\tilde{\bs{\mu}},\tilde{\bs{\Sigma}},g)$. Specifically, (see \cite[Sec. 3.5]{RichPhD} and \cite[Sec. 4.2.2]{book_zoubir}):
\be
\label{RES_CES}
\begin{split}
	R&ES_{2N}(\tilde{\mb{x}};\tilde{\bs{\mu}},\tilde{\bs{\Sigma}},g) \triangleq p_{\tilde{X}}(\tilde{\mb{x}};\tilde{\bs{\mu}},\tilde{\bs{\Sigma}},g)\\
	&=2^{-(2N)/2}|\mb{\tilde{\bs{\Sigma}}}|^{-1/2} g ((\tilde{\mb{x}}-\tilde{\bs{\mu}})^T \tilde{\bs{\Sigma}}^{-1}(\tilde{\mb{x}}-\tilde{\bs{\mu}})^T )\\
	&=|\bs{\Sigma}|^{-1} g \left(2(\mb{z}-\bs{\mu})^H \bs{\Sigma}^{-1}(\mb{z}-\bs{\mu}) \right)\\
	&=p_{Z}(\mb{z};\bs{\mu},\bs{\Sigma},h) \triangleq CES_{N}(\mb{z};\bs{\mu},\bs{\Sigma},h),
\end{split}
\ee 
where $h(t) \triangleq g(2t)$. Note that by moving from the real to the complex representation, the functional form of the density generator remains unchanged except for the scaling factor 2 of its argument. Furthermore, the pdf of a CES-distributed random vector $\mb{z}$ can be expressed as\footnote{Note that this definition is consistent with the one proposed in \cite{Esa} except for the normalizing constant $c_{N,g}$ that we included in the functional form of density generator $h$.}:
\be
\label{true_CES}
\begin{split}
	p_{Z}(\mb{z}|\bs{\theta}, h) 
	=|\bs{\Sigma}|^{-1} h \left(  (\mb{z}-\bs{\mu})^{H}
	\bs{\Sigma}^{-1}(\mb{z}-\bs{\mu}) \right).
\end{split}
\ee

As for RES distributed vectors, any CES distributed vector $\mb{z}$ can be represented as (\cite{CES_stat}, \cite{Esa} and \cite[Sec. 3.5]{RichPhD}):
\be
\label{CSRT_dec}
\mb{z} =_d \bs{\mu} + \sqrt{\mathcal{Q}}\bs{\Sigma}^{1/2}\mb{u},
\ee
where $\mb{u} \sim \mathcal{U}(\mathbb{C}S^N)$ is a complex random vector uniformly distributed on the unit complex $N$-sphere $\mathbb{C}S^N$ and $\mathcal{Q}$ is the so-called \textit{2nd-order modular variate}, s.t.:
\be
\label{somv}
\mathcal{Q} =_d Q \triangleq (\mb{z}-\bs{\mu})^H \bs{\Sigma}^{-1}(\mb{z}-\bs{\mu}),
\ee
whose pdf is given by:
\be
\label{CSS_Q_pdf}
p_{\mathcal{Q}}(q) = 2^{-1} s_N q^{N-1} h \left(q \right) = \pi^{N}\Gamma(N)^{-1}q^{N-1} g \left(2q \right),
\ee
where $s_N \triangleq 2\pi^{N}/\Gamma(N)$ is the surface area of $\mathbb{C}S^N$. From \eqref{CSRT_dec} and by exploiting the properties of $\mb{u}$ \cite[Lemma 1]{Esa}, we have that the covariance matrix of the CES-distributed vector $\mb{z}$ is $\mb{M} \triangleq E\{(\mb{z}-\bs{\mu})(\mb{z}-\bs{\mu})^H\} = N^{-1}E\{\mathcal{Q}\}\mb{\Sigma}$.    

It is immediate to verify that the representation in \eqref{CSRT_dec} is scale-ambiguous since $\mb{z} =_{d} \bs{\mu }+\sqrt{\mathcal{Q}}\bs{\Sigma}^{-1/2}\mb{u} =_{d} \bs{\mu}+\sqrt{c^{-2}\mathcal{Q}}(c\bs{\Sigma}^{-1/2})\mb{z}, \forall c>0$. Moreover, as for the RES case, the scale ambiguity appears also in the functional representation of a CES pdf since $CES_{N}(\mb{z};\bs{\mu},\bs{\Sigma},h(t))\equiv CES_{N}(\mb{z};\bs{\mu},c\bs{\Sigma},h(ct)), \forall c>0$. There are two different, yet equivalent, ways to avoid this scale ambiguity. The first one is to put a constraint on the scatter matrix $\bs{\Sigma}$, e.g. we may choose to impose the usual constraint on its trace as done in \cite{For_SCRB}, that is $\mathrm{tr}(\bs{\Sigma})=N$. The second equivalent approach is to impose a constraint on the functional form of the density generator $h$. Following the same procedure adopted in \cite{miss_sb}, we may assume that $h \in \mathcal{G}$ is parameterized in order to satisfy the constraint:
\be
\label{const}
E\{\mathcal{Q}\}=\pi^{N}\Gamma(N)^{-1}\int_{0}^{+\infty}q^{N-1} h (q)dq=N.
\ee
As a consequence of \eqref{const}, the scatter matrix $\bs{\Sigma}$ equates the covariance matrix $\mb{M}$ of $\mb{z}$ \cite[Sec. III.C]{Esa}. For further reference, we define the set $\bar{\mathcal{G}} \subset \mathcal{G}$ as the set of all the density generators satisfying the constraint in \eqref{const}. Moreover, all the expectation operators taken w.r.t. the \virg{constrained} pdf of the second-order modular variate in \eqref{somv} will be indicated as $\bar{E}\{\cdot\}$, s.t.
\be
\label{bar_E}
\bar{E}\{f(\mathcal{Q})\} \triangleq \int_{0}^{+\infty}f(q) p_{\mathcal{Q}}(q)dq = \pi^{N}\Gamma(N)^{-1}\int_{0}^{+\infty}f(q)q^{N-1} h (q)dq, \; h \in \bar{\mathcal{G}}.
\ee

As we will discuss ahead in the paper, in order to obtain the constrained SCRB on the joint estimation of $\bs{\mu}$ and $\bs{\Sigma}$, we will exploit the constraint on the trace of $\bs{\Sigma}$, while to derive the SSB formula we will rely on the constraint on the desity generator given in $\eqref{const}$.

Definition \ref{def_CES} and the equality chain in \eqref{RES_CES} suggest the existence of a one-to-one mapping between the subset of the RES distributions satisfying the covariance structure specified in \eqref{scatter_real} and the family of CES distributions. In other words, the CES \virg{framework} is just a convenient and compact representation of a \textit{subset} of RES distributions. This implies that the theory already developed for the RES class holds true for the CES class as well. In particular, by relying on the approach proposed in \cite[Sec. 3.5]{RichPhD}, CES distributions can be interpreted as the \textit{semiparametric group model} generated by the set of Complex Spherically Symmetric (CSS) distributions through the action of the group of affine transformations:
\be
\begin{split}
	\alpha_{(\bs{\mu},\bs{\Sigma})}: \; & \mathbb{C}^N \rightarrow \mathbb{C}^N, \; \forall \bs{\mu}, \bs{\Sigma} \\
	 & \mb{z}\mapsto \alpha_{(\bs{\mu},\bs{\Sigma})}(\mb{z}) = \bs{\mu} + \bs{\Sigma}^{1/2}\mb{z}.
\end{split}
\ee
Then, the semiparametric structure detailed in \cite[Sec. 3]{For_SCRB} for the RES distribution can be directly translated in the CES context without any new specific manipulations.

\section{The constrained SCRB for complex parameter estimation in CES distributions}
\label{CRB_sec}
In this section, a closed form expression of the constrained CSCRB for the joint estimation of the complex mean vector $\bs{\mu}$ and of the complex constrained scatter matrix $\bs{\Sigma}$ of CES-distributed vectors is provided. The subsequent derivation strictly follows the one described in \cite{For_SCRB} for the real case. However, in the complex case, the derivatives have to be considered as \textit{Wirtinger derivatives}. More precisely, following Theorem IV.1 in \cite{For_SCRB}, the steps are:
\begin{itemize}
	\item[\textit{A}.] Define the complex constrained parameter space $\bar{\Omega}_\mathbb{C}$.
	\item[\textit{B}.] Evaluate the semiparametric efficient score vector $\bar{\mb{s}}_0(\mb{z})$ using the Wirtinger derivatives.
	\item[\textit{C}.] Derive the SFIM for the joint estimation of $\bs{\mu}$ and $\bs{\Sigma}$.
	\item[\textit{D}.] Obtain a closed form expression for the complex CSCRB.
\end{itemize}

\subsection{The complex constrained parameter space $\bar{\Omega}_\mathbb{C}$}
As mentioned before, the parametric part of the semiparametric CES model is given by the mean vector $\bs{\mu}$ and by the Hermitian scatter matrix $\bs{\Sigma}$. According to the rules of the Wirtinger calculus, to define a complex parameter space, we have to take into account the parameters to be estimated together with their complex conjugates \cite{Bos,Rao,OllCRB,Menni, Rich, Complex_MCRB_For}. To this end, we note that, while $\bs{\mu}$ is composed of $N$ complex free parameters, i.e. all its $N$ entries, the Hermitian scatter matrix $\bs{\Sigma}$ can be parametrized by means of its $N$ real diagonal entries and of its $N(N-1)/2$ complex entries that are positioned strictly below the main diagonal \cite{Complex_M}. More formally, and following the notation in \cite{Menni} and \cite{Rich}, the parametric part of the CES model can be described by the parameter vector $\bs{\theta} = (\bs{\theta}_c^T,\bs{\theta}_c^H,\bs{\theta}_r^T)^T$, where:
\be
\label{def_theta}
\bs{\theta}_c = (\bs{\mu}^T,\vecl{\bs{\Sigma}}^T)^T, \quad \bs{\theta}_r = \mathrm{diag}(\bs{\Sigma}),
\ee
the operator $\vecl{\cdot}$ selects all the entries strictly below the main diagonal of $\bs{\Sigma}$ taken in the same column-wise order as the ordinary $\mathrm{vec}(\cdot)$ operator \cite[Sec. 2.4]{Complex_M} while $\mathrm{diag}(\bs{\Sigma})$ is a column vector collecting the diagonal elements of $\bs{\Sigma}$.

For ease of calculation, we express the parameter vector $\bs{\theta}$ with respect to a different basis. In particular, let us introduce a permutation matrix $\mb{P}$, s.t.:
\be
\label{def_phi}
\bs{\phi} \triangleq (\bs{\mu}^T,\bs{\mu}^H,\vcsigma^T)^T= \mb{P} \bs{\theta}.
\ee
It is worth stressing here that the previous two characterizations of the \textit{augmented} complex parameter vectors $\bs{\theta}$ and $\bs{\phi}$ given in \eqref{def_theta} and \eqref{def_phi} are equivalent, since the scatter matrix $\bs{\Sigma}$ is an Hermitian matrix and the permutation matrix $\mb{P}$ only represents an orthogonal change of basis \cite[Sec. 6.5.5]{Complex_M}. Consequently, let us define the \virg{augmented} complex parameter space $\Omega_\mathbb{C} \subset \mathbb{C}^q$ of dimension $q=N(N+2)$ as:
\be
\label{par_space}
\Omega_\mathbb{C} = \{\bs{\phi} \in \mathbb{C}^q | \bs{\phi}\; \mathrm{is \; as \; in \; \eqref{def_phi}};\bs{\mu} \in \mathbb{C}^N,\bs{\Sigma} \in \mathcal{M}_N^{\mathbb{C}}\},
\ee
where $\mathcal{M}_N^{\mathbb{C}}$ is the set of all the Hermitian, positive-definite matrices of dimension $N \times N$. As previously discussed, in order to avoid the scale ambiguity between the scatter matrix and the density generator of a CES distribution, we choose to impose a constraint on the trace of $\bs{\Sigma}$. Specifically, let us define the scalar, real-valued, constraint function as:
\be
\label{const_function}
c(\bs{\Sigma}) \triangleq \mathrm{tr}(\bs{\Sigma})-N =0.
\ee
Then, the function $c(\bs{\Sigma})$ constrains the parameter vector $\bs{\phi}$ in a smooth sub-manifold of $\Omega_\mathbb{C}$ defined as:
\be
\label{Omega_bar}
\bar{\Omega}_\mathbb{C} = \{\bs{\phi} \in \Omega_\mathbb{C} |c(\bs{\Sigma})=0\},
\ee
of dimension $\bar{q}=q-1$. From now on, $\bar{\Omega}_\mathbb{C}$ will be considered as the reference parameter space.

\subsection{The complex semiparametric efficient score vector $\bar{\mb{s}}_0(\mb{z})$}
This subsection provides a closed form expression for the semiparametric efficient score vector $\bar{\mb{s}}_0 \equiv \bar{\mb{s}}_0(\mb{z})$, evaluated at the true parameter vector $\bs{\phi}_0 \in \Omega_{\mathbb{C}}$. The complex extension of the semiparametric efficient score vector given in \cite[Theo IV.1]{For_SCRB} can be defined as:
\be
\label{eff_com_scor_vect}
\bar{\mb{s}}_0 = [\bar{\mb{s}}^T_{\bs{\mu}_0},\bar{\mb{s}}^T_{\bs{\mu}^*_0},\bar{\mb{s}}^T_{\mathrm{vec}(\bs{\Sigma}_0)}]^T = \mb{s}_{\bs{\phi}_0} - \Pi(\mb{s}_{\bs{\phi}_0}|\mathcal{T}_{h_0}),
\ee 
where $\mb{s}_{\bs{\phi}_0}$ is the score vector w.r.t. $\bs{\phi}_0$ and $\Pi(\mb{s}_{\bs{\phi}_0}|\mathcal{T}_{h_0})$ is the orthogonal projection of $\mb{s}_{\bs{\phi}_0}$ on the nuisance tangent space evaluated at the true density generator $h_0$.

The score vector w.r.t. $\bs{\phi}_0$ can be expressed as:
\be
\mb{s}_{\bs{\phi}_0} \triangleq \nabla_{\bs{\phi}} \ln p_Z(\mb{z};\bs{\phi}_0,h_0) = [\mb{s}^T_{\bs{\mu}_0},\mb{s}^T_{\bs{\mu}^*_0},\mb{s}^T_{\vsigmat}]^T
\ee
where, following the approach detailed in \cite{Menni}, the complex gradient operator of a scalar, real-valued, function $f(\bs{\phi})$, evaluated in $\bs{\phi}_0$, is defined as:
\be
[\nabla_{\bs{\phi}} f(\bs{\phi}_0)]_i = \left. \partial f(\bs{\phi})/ \partial\phi_i^* \right|_{\bs{\phi}=\bs{\phi}_0}, \; i=1,\ldots,q .
\ee

The closed form expression for $\mb{s}_{\bs{\mu}_0}$, $\mb{s}_{\bs{\mu}^*_0}$ and $\mb{s}^T_{\vsigmat}$ can be obtained by applying the standard rules of the Wirtiger matrix calculus. For an excellent and comprehensive book about this topic, we refer the reader to \cite{Complex_M}. Here, to not clutter the presentation with too many technicalities, we will provide only the final outcomes without reporting all the steps.     

The complex gradient w.r.t. $\bs{\mu}$ of $\ln p_Z(\mb{z};\bs{\phi}_0,h_0)$ can be obtained by applying the rules listed in Table 4.2 of \cite{Complex_M} as:
\be
\label{s_mu}
\begin{split}
	\mb{s}_{\bs{\mu}_0}(\mb{z}) &= - \psi_0(Q_0)\bs{\Sigma}_0^{-1}(\mb{z}-\bs{\mu}_0)=_d -\sqrt{\mathcal{Q}}\psi_0(\mathcal{Q})\bs{\Sigma}_0^{-1/2}\mb{u}.
\end{split}
\ee
Consequently, we have that:
\be
\label{s_mu_star}
\mb{s}_{\bs{\mu}^*_0}(\mb{z}) = \mb{s}^*_{\bs{\mu}_0}(\mb{z}) =_d -\sqrt{\mathcal{Q}}\psi_0(\mathcal{Q})(\bs{\Sigma}_0^*)^{-1/2}\mb{u}^*,
\ee
where 
\be
\label{psi}
\psi_0(t) \triangleq d \ln h_0(t)/dt.
\ee
Moreover, by applying the derivative rules listed in Table 4.3 and the equality in \cite[eq. 6.199]{Complex_M}, we get: 
\be
\label{s_vec}
\begin{split}
	\mb{s}_{\mathrm{vec}(\bs{\Sigma}_0)}(\mb{z}) &= -\mathrm{vec}(\bs{\Sigma}_0^{-1}) - \psi_0(Q_0) \bs{\Sigma}_0^{-*} \otimes \bs{\Sigma}_0^{-1} \mathrm{vec}\left( (\mb{z}-\bs{\mu}_0)(\mb{z}-\bs{\mu}_0)^H \right)\\
	&=_d -\mathrm{vec}(\bs{\Sigma}_0^{-1}) - \mathcal{Q}\psi_0(\mathcal{Q}) ( (\bs{\Sigma}_0^*)^{-1/2} \otimes \bs{\Sigma}_0^{-1/2} )  \mathrm{vec}(\mb{u}\mb{u}^H).
\end{split}
\ee

The next step is the derivation of the orthogonal projection of the score vector $\mb{s}_{\bs{\phi}_0}$ on the nuisance tangent space of the CES semiparametric group model evaluated at the true density generator $h_0$. The procedure to obtain a closed form expression for $\Pi(\mb{s}_{\bs{\phi}_0}|\mathcal{T}_{h_0})$ parallels the one described in \cite[Sec. IV.B]{For_SCRB} for the real case. Specifically, the properties of the semiparametric group models collected in Proposition II.1 of \cite{For_SCRB} can be applied to derive $\Pi(\mb{s}_{\bs{\phi}_0}|\mathcal{T}_{h_0})$. Then, by replicating step-by-step the procedure discussed in \cite[Sec. IV.B]{For_SCRB}, we obtain:  
\be
\label{proj_mu}
\Pi(\mb{s}_{\bs{\mu}_0}|\mathcal{T}_{h_0}) = \Pi(\mb{s}_{\bs{\mu}^*_0}|\mathcal{T}_{h_0}) = \mb{0}_N,
\ee
\be
\label{proj_vect}
\begin{split}
	\Pi(\mb{s}_{\mathrm{vec}(\bs{\Sigma}_0)}&|\mathcal{T}_{h_0}) = -(1+N^{-1}\mathcal{Q}\psi_0(\mathcal{Q}))\mathrm{vec}(\bs{\Sigma}_0^{-1}).
\end{split}
\ee
Note that, as for the real case, $\mb{s}_{\bs{\mu}_0}$ and $\mb{s}_{\bs{\mu}^*_0}$ are orthogonal to the nuisance tangent space $\mathcal{T}_{h_0}$. This implies that we achieve the same (asymptotic) performance in the estimation of $\bs{\mu}_0$ by knowing or not knowing the true density generator $h_0$.

The efficient score vector $\bar{\mb{s}}_0$ in \eqref{eff_com_scor_vect} can now be derived by collecting previous results. In particular, we have that $\bar{\mb{s}}_{\bs{\mu}_0} \equiv \mb{s}_{\bs{\mu}_0}$ and $\bar{\mb{s}}_{\bs{\mu}^*_0} \equiv \mb{s}_{\bs{\mu}^*_0}$ since, as reported in \eqref{proj_mu}, the projection is nil, and
\be
\label{eff_score_vect}
\begin{split}
	\bar{\mb{s}}_{\mathrm{vec}(\bs{\Sigma}_0)}=_d \mathcal{Q}\psi_0(\mathcal{Q})& ( (\bs{\Sigma}_0^*)^{-1/2} \otimes \bs{\Sigma}_0^{-1/2} \mathrm{vec}(\mb{u}\mb{u}^H) - N^{-1}\mathrm{vec}(\bs{\Sigma}_0^{-1}) ) . 
\end{split}
\ee
\subsection{The SFIM $\bar{\mb{I}}(\bs{\phi}_0|h_0)$}
The SFIM can be expressed as the following block matrix:
%\be
%\label{SFIM_com}
%\bar{\mb{I}}(\bs{\phi}_0|\bar{g}_0) = \left(
%\begin{array}{cc}
%	\begin{array}{cc}
%		\mb{C}_0(\bar{\mb{s}}_{\bs{\mu}_0}) & \mb{0}_{N \times N}\\
%		\mb{0}_{N \times N} & \mb{C}_0^*(\bar{\mb{s}}_{\bs{\mu}_0})
%	\end{array} & \mb{0}_{2N \times N^2}\\
%	\mb{0}_{2N \times N^2} & \mb{C}_0(\bar{\mb{s}}_{\mathrm{vec}(\bs{\Sigma}_0)})
%\end{array}
%\right)
%\ee
\be
\label{SFIM_com}
\bar{\mb{I}}(\bs{\phi}_0|h_0) = \left(
\begin{array}{cc}
	\bar{\mb{I}}(\bs{\mu}_0|h_0) & \mb{0}_{2N \times N^2}\\
	\mb{0}_{N^2 \times 2N} & \mb{C}_0(\bar{\mb{s}}_{\mathrm{vec}(\bs{\Sigma}_0)})
\end{array}
\right),
\ee
where, for a generic function $\mb{l} \equiv \mb{l}(\mb{z})$, we define $\mb{C}_0(\mb{l})\triangleq E_0\{\mb{l}\mb{l}^H\}$. The off-diagonal block matrices in \eqref{SFIM_com} vanish because all the third-order moments of $\mb{u}$ vanish \cite[Lemma 1]{Esa} and
\be
\label{SFIM_com_mu}
\bar{\mb{I}}(\bs{\mu}_0|h_0) =	\left( \begin{array}{cc}
	\mb{C}_0(\bar{\mb{s}}_{\bs{\mu}_0}) & \mb{0}_{N \times N}\\
	\mb{0}_{N \times N} & \mb{C}_0^*(\bar{\mb{s}}_{\bs{\mu}_0}),
\end{array}\right), 
\ee
\be
\label{FIM_mu}
\mb{C}_0(\bar{\mb{s}}_{\bs{\mu}_0}) = N^{-1}E\{\mathcal{Q}\psi_0(\mathcal{Q})^2\}\bs{\Sigma}_0^{-1}.
\ee
Note that the off-diagonal matrices in \eqref{SFIM_com_mu} vanish due to the circularity of $\mb{u}$, while to derive \eqref{FIM_mu}, we used the fact that $E\{\mb{u}\mb{u}^H\} = N^{-1}\mb{I}$ \cite[Lemma 1]{Esa}. Moreover, after some standard complex matrix manipulations, we get:
\be
\label{cov_mat_eff_scat_com}
\begin{split}
	\mb{C}_0&(\bar{\mb{s}}_{\mathrm{vec}(\bs{\Sigma}_0)}) = \frac{E\{\mathcal{Q}^2\psi_0(\mathcal{Q})^2\}}{N(N+1)}  ( \kronsigmatinvT - N^{-1}\vcsigmatinv \vcsigmatinv^H ).
\end{split}
\ee
It is worth noticing that the constraint on the trace of the scatter matrix $\bs{\Sigma}_0$ has not been imposed yet. 
\subsection{The complex constrained SCRB: $\mathrm{CCSCRB}(\bs{\phi}_0|h_0)$}
\label{const_SCRB}
We are now ready to derive a closed form expression of the SCRB for the constrained estimation of the complex parameter vector $\bs{\phi}_0 \in \bar{\Omega}_{\mathbb{C}}$, i.e. $\mathrm{CCSCRB}(\bs{\phi}_0|h_0)$. As showed in \cite[Theo. IV.1]{For_SCRB} for the real case, the first step to obtain $\mathrm{CCSCRB}(\bs{\phi}_0|h_0)$ is the derivation of the matrix $\mb{U}$ whose columns form an orthonormal basis for the null space of the Jacobian matrix of the constraint function $c(\bs{\Sigma}_0)$ in \eqref{const_function}. Since, in our case, $c(\bs{\Sigma}_0)$ involves only the \textit{real} diagonal elements of the Hermitian matrix $\bs{\Sigma}_0$, $\mb{U} \in \mathbb{R}^{N^2 \times (N^2-1)}$ is the matrix that satisfies the following two conditions:
\be
\label{U_prop}
\nabla^T_{\mathrm{vec}(\bs{\Sigma})}c(\bs{\Sigma}_0)\mb{U}=\mb{0}, \qquad \mb{U}^T\mb{U}=\mb{I}_{N^2-1}.
\ee
Through direct calculation, we have that:
\be
\nabla^T_{\mathrm{vec}(\bs{\Sigma})}c(\bs{\Sigma}_0) = \mathrm{vec}(\mb{I}_{N})^T.
\ee
Then, matrix $\mb{U}$ can be obtained numerically by evaluating the $N^2-1$ orthonormal eigenvectors associated with the zero eigenvalue of $\mathrm{vec}(\mb{I}_{N})^T$ through SVD.

Finally, the CCSCRB for the estimation of $\bs{\phi}_0 \in \bar{\Omega}_{\mathbb{C}}$ in \eqref{Omega_bar} can be expressed as:
\be
\label{SCRB_gun_com}
\mathrm{CCSCRB}(\bs{\phi}_0|h_0) = \left( \begin{array}{cc}
	\bar{\mb{I}}(\bs{\mu}_0|h_0)^{-1} & \mb{0}_{2N \times N^2} \\
	\mb{0}_{N^2 \times 2N}   & 	\bar{\mb{I}}(\bs{\Sigma}_0|h_0)^{-1}
\end{array}\right),
\ee
where the two block-diagonal matrices are the inverse of the SFIMs for the estimation of the mean vector $\bs{\mu}_0$ and of the constrained scatter matrix $\bs{\Sigma}_0$ that can be expressed as:
\be
\label{inv_SFIM_com_mu}
\bar{\mb{I}}(\bs{\mu}_0|h_0)^{-1} = \frac{N}{E\{\mathcal{Q}\psi_0(\mathcal{Q})^2\}}	\left( \begin{array}{cc}
	\bs{\Sigma}_0 & \mb{0}_{N \times N}\\
	\mb{0}_{N \times N} & \bs{\Sigma}_0^*
\end{array}\right),
\ee
\be
\label{inv_SFIM_com_Sigma}
\bar{\mb{I}}(\bs{\Sigma}_0|h_0)^{-1} = \mb{U}\left( \mb{U}^T\mb{C}_0(\bar{\mb{s}}_{\vsigmat})\mb{U}\right)^{-1}\mb{U}^T.
\ee
Note that, as for the real case, the block-diagonal structure of $\mathrm{CCSCRB}(\bs{\phi}_0|h_0)$ implies that not knowing the mean vector $\bs{\mu}_0$ have no impact on the optimal \textit{asymptotic} performance in the estimation of the scatter matrix $\bs{\Sigma}_0$. A numerical example of the calculation of the $\mathrm{CCSCRB}(\bs{\phi}_0|h_0)$ in complex $t$-distributed data will be given in Sect. \ref{num_t_CCSCRB}, where the efficiency of two scatter matrix estimators is investigated through simulations. 

\section{Semiparametric Slepian-Bangs formula for CES distributions}
\label{SSB_sec}
Eqs. \eqref{SCRB_gun_com}, \eqref{inv_SFIM_com_mu} and \eqref{inv_SFIM_com_Sigma} provide a closed form expression for the CSCRB for the joint estimation of the mean vector and the scatter matrix of a CES-distributed vector. In this section, we focus our attention on a more general case where both the mean vector and the scatter matrix can be parametrized by a \textit{real} parameter vector. Let us start with some preliminaries. Let $\mathbb{C}^N \ni \mb{z} \sim CES_N(\bs{\mu}(\bs{\theta}),\bs{\Sigma}(\bs{\theta}), h)$ be a CES-distributed random vector whose mean value $\bs{\mu}(\bs{\theta}) \in \mathbb{C}^N$ and scatter matrix $\bs{\Sigma}(\bs{\theta}) \in \mathbb{C}^{N \times N}$ are parameterized by a $d$-dimensional parameter vector $\bs{\theta} \in \Theta \subset \mathbb{R}^d$ to be estimated. The density generator $h \in \bar{\mathcal{G}}$ is left unspecified since it represents an unknown, infinite-dimensional nuisance parameter. We assume here that $\bs{\Sigma}(\bs{\theta})$ is a full rank, positive definite, Hermitian matrix for any possible value of $\bs{\theta} \in \Theta$. Consequently, the pdf of $\mb{z}$ can be expressed as shown in \eqref{true_CES}.

To avoid the scale ambiguity problem between the scatter matrix and the density generator, we impose the constraint in \eqref{const} on the functional form of $h$. We were steered towards this choice just by the ease of calculation. Here, in fact, the scatter matrix is parametrized by the vector of interest $\bs{\theta}$ and it is not easy to work with a constrained parametric scatter matrix. The adoption of the constraint on $h$ leaves $\bs{\Sigma}(\bs{\theta})$ unconstrained and this greatly simplifies the derivation. Note that, as a consequence of \eqref{const}, the scatter matrix $\bs{\Sigma}(\bs{\theta})$ is the covariance matrix of $\mb{z}$.
 
We now focus our attention on the \textit{semiparametric group} nature of the family of all the pdfs, say $\mathcal{P}_{\bs{\theta},h}$, of an (absolutely continuous) CES-distributed random vector $\mb{z} \sim CES_N(\bs{\mu}(\bs{\theta}),\bs{\Sigma}(\bs{\theta}), h)$ with $\bs{\theta} \in \Theta$ and $h \in \bar{\mathcal{G}}$. Following the discussion provided in \cite[Sec. 4.2 and 4.3]{BKRW}, let us firstly introduce the group $\mathcal{A}$ of affine transformations:
\be
\label{affine_tran}
\begin{split}
	\mathcal{A} \ni \alpha_{\bs{\theta}}: \; & \mathbb{C}^N \rightarrow \mathbb{C}^N, \; \forall \bs{\theta} \in \Theta \\
	\mathbb{C}^N \ni & \mb{w}\mapsto \alpha_{\bs{\theta}}(\mb{w})= \bs{\mu}(\bs{\theta}) + \bs{\Sigma}(\bs{\theta})^{1/2}\mb{w}.
\end{split}
\ee
Then, as shown in \cite[Sec. 4.2, Lemma 2]{BKRW}, the model $\mathcal{P}_{\bs{\theta},h}$ can be considered as a \textit{semiparametric group model} generated by $\mathcal{A}$ and it can be explicitly expressed as:
\be
\label{CES_semi_group_model}
\mathcal{P}_{\bs{\theta},h} = \lbr p_Z | p_Z(\mb{z}|\bs{\theta},h) = |\bs{\Sigma}(\bs{\theta})|^{-1} h(\norm{\alpha^{-1}_{\bs{\theta}}(\mb{z})}^2),  \bs{\theta} \in \Theta, h \in \bar{\mathcal{G}} \rbr,
\ee 
where $\alpha^{-1}_{\bs{\theta}}(\cdot)= \bs{\Sigma}(\bs{\theta})^{-1/2}(\cdot-\bs{\mu}(\bs{\theta}))$ is the inverse transformation of $\alpha_{\bs{\theta}} \in \mathcal{A}$ and $\norm{\cdot}$ indicates the Euclidean norm. Under some regularity conditions on the mapping $\bs{\theta}\rightarrow(\bs{\mu}(\bs{\theta}),\bs{\Sigma}(\bs{\theta}))$ discussed in \cite[Sec. 4.2, pp. 92, Assumptions (iii), (iv), (v)]{BKRW}, we can exploit the properties of the semiparametric group models to evaluate the Semiparametric FIM $\bar{\mb{I}}(\bs{\theta}_0|h_0)$ for the estimation of the \virg{true} parameter vector $\bs{\theta}_0 \in \Theta$ in the presence of the \virg{true} nuisance density generator $h_0 \in \bar{\mathcal{G}}$. The closed form expression for $\bar{\mb{I}}(\bs{\theta}_0|h_0)$ that we are going to derive is exactly the SSB formula.  

\subsection{The single snapshot case}
\label{single_snap}
Let us start with the case in which we have only one \textit{snapshot} sampled from an unspecified CES distribution, i.e. $\mb{z} \sim CES_N(\bs{\mu}_0,\bs{\Sigma}_0, h_0)$, where $\bs{\mu}_0 \equiv \bs{\mu}(\bs{\theta}_0)$ and $\bs{\Sigma}_0 \equiv \bs{\Sigma}(\bs{\theta}_0)$. As discussed in \cite[Sec. 3.4]{BKRW} and recalled in \cite[Sec. IV.B]{For_SCRB}, the SFIM for the estimation of $\bs{\theta}_0 \in \Theta$ is defined as $\bar{\mb{I}}(\bs{\theta}_0|h_0)\triangleq E_0\{\bar{\mb{s}}_{\bs{\theta}_0}\bar{\mb{s}}_{\bs{\theta}_0}^{H}\}$ where the semiparametric efficient score vector $\bar{\mb{s}}_{\bs{\theta}_0}\equiv \bar{\mb{s}}_{\bs{\theta}_0}(\mb{z})$ is given by:
\be
\label{semi_eff_score}
\bar{\mb{s}}_{\bs{\theta}_0} \triangleq  \mb{s}_{\bs{\theta}_0} - \Pi(\mb{s}_{\bs{\theta}_0}|\mathcal{T}_{h_0}),
\ee
where $\mb{s}_{\bs{\theta}_0}$ is the score vector evaluated at the true parameter vector $\bs{\theta}_0$ and $\Pi(\mb{s}_{\bs{\theta}_0}|\mathcal{T}_{h_0})$ is the orthogonal projection of $\mb{s}_{\bs{\theta}_0}$ on the semiparametric nuisance tangent space $\mathcal{T}_{h_0}$ of $\mathcal{P}_{\bs{\theta},h}$ in \eqref{CES_semi_group_model} evaluated at the true density generator $h_0$. The procedure that we have to follow in order to obtain the SSB formula, i.e. the closed form expression of $\bar{\mb{I}}(\bs{\theta}_0|h_0)$ is similar to the one adopted in Sec. \ref{CRB_sec} to derive the CSCRB for the joint estimation of $\bs{\mu}$ and $\bs{\Sigma}$:
\begin{enumerate}
\item  Evaluate the semiparametric score vector $\bar{\mb{s}}_{\bs{\theta}_0}$.
\item  Calculate the SFIM $\bar{\mb{I}}(\bs{\theta}_0|h_0)\triangleq E_0\{\bar{\mb{s}}_{\bs{\theta}_0}\bar{\mb{s}}_{\bs{\theta}_0}^{H}\}$ 
\item Rearrange the $\bar{\mb{I}}(\bs{\theta}_0|h_0)$ in a compact and easy-to-use expression, i.e. the SSB formula. 
\end{enumerate}    

In the following, the above-mentioned three steps are developed in details.

\subsubsection{Evaluation of the semiparametric efficient score vector $\bar{\mb{s}}_{\bs{\theta}_0}$}
Let us start with the calculation of the score function $\mb{s}_{\bs{\theta}_0}$. Following the derivation in \cite[Sec. 3.1]{miss_sb} and \cite[Sec. III]{Bess}, each entry of $\mb{s}_{\bs{\theta}_0}$ can be easily evaluated as:
\be
\label{grad_f}
\begin{split}
	[\mb{s}_{\bs{\theta}_0}]_i & \triangleq \left. \parder{\ln p_{Z}\left( \mb{z};\bs{\theta}\right) }{\theta_i}\right|_{\bs{\theta}=\bs{\theta}_0} = \mathrm{tr}(\mb{P}_i^0) + \psi_0(Q_0)\parder{Q_0}{\theta_i},
\end{split}
\ee
where the function $\psi_0$ has already been defined in \eqref{psi} and $\mb{P}_i^0 \triangleq \bs{\Sigma}_0^{-1/2}\bs{\Sigma}_i^0\bs{\Sigma}_0^{-1/2}$. Moreover, from \cite[eq. (22)]{miss_sb} and \cite[eq. (8)]{Bess}, we have:
\be
\label{dev_Q_0}
\parder{Q_0}{\theta_i} =-2\mathrm{Re}\left[ (\mb{z} - \bs{\mu}_0)^H\bs{\Sigma}_0^{-1} \bs{\mu}_i^0 \right] -(\mb{z} - \bs{\mu}_0)^H\mb{S}_i^0(\mb{z} - \bs{\mu}_0),
\ee 
where, according to the notation previously introduced, $\bs{\mu}_i^0 \triangleq \parder{ \bs{\mu}_0}{\theta_i}$ and $\mb{S}_i^0=\bs{\Sigma}_0^{-1}\bs{\Sigma}_i^0\bs{\Sigma}_0^{-1}$. By collecting previous results, the entries of the score vector $\mb{s}_{\bs{\theta}_0}$ can be expressed as:
\be
\label{score_vect_SB}
\begin{split}
	[\mb{s}_{\bs{\theta}_0}]_i & = \mathrm{tr}\left(\mb{P}_i^0\right) - \psi_0(Q_0) \left( 2\mathrm{Re}\left[ (\mb{z} - \bs{\mu}_0)^H\bs{\Sigma}_0^{-1} \bs{\mu}_i^0 \right]+\right. \\ 
	& \left. +(\mb{z} - \bs{\mu}_0)^H\mb{S}_i^0(\mb{z} - \bs{\mu}_0) \right), \; i=1,\ldots,d. 
\end{split}
\ee
Using the representation given in \eqref{CSRT_dec}, eq. \eqref{score_vect_SB} can be rewritten as:
\be
\label{score_vect_rep}
\begin{split}
	[\mb{s}_{\bs{\theta}_0}]_i &=_d  - \psi_0(\mathcal{Q}) \left( 2\sqrt{\mathcal{Q}}\mathrm{Re}\left[ \mb{u}^H \bs{\Sigma}_0^{H/2}\bs{\Sigma}_0^{-1} \bs{\mu}_i^0 \right] + \mathcal{Q}\mb{u}^H \bs{\Sigma}_0^{H/2}\mb{S}_i^0\bs{\Sigma}_0^{1/2}\mb{u} \right) +\mathrm{tr}\left(\mb{P}_i^0\right) \\
	&= - \psi_0(\mathcal{Q}) \left( 2\sqrt{\mathcal{Q}}\mathrm{Re}\left[ \mb{u}^H \bs{\Sigma}_0^{-1/2}\bs{\mu}_i^0 \right]  +\mathcal{Q}\mb{u}^H \mb{P}_i^0 \mb{u} \right) + \mathrm{tr}\left( \mb{P}_i^0 \right), \; i=1,\ldots,d. 
\end{split}
\ee

The orthogonal projection $\Pi(\mb{s}_{\bs{\theta}_0}|\mathcal{T}_{h_0})$ can be obtained by following exactly the same procedure discussed in \cite[Sec. IV.B]{For_SCRB}. For the sake of conciseness, here we report only the final result as:
\be
\label{proj_ope}
\begin{split}
	[\Pi(&\mb{s}_{\bs{\theta}_0}|\mathcal{T}_{h_0})]_{i} = E_{0|\sqrt{\mathcal{Q}}}\{[\mb{s}_{\bs{\theta}_0}]_{i}|\sqrt{\mathcal{Q}}\}\\
	& =_d \mathrm{tr}( \mb{P}_i^0)- 2\sqrt{\mathcal{Q}}\psi_0(\mathcal{Q})\mathrm{Re}\left[ E\{\mb{u}\}^H \bs{\Sigma}_0^{-1/2}\bs{\mu}_i^0 \right] - \mathcal{Q} \psi_0(\mathcal{Q}) \mathrm{tr}\left(\mb{P}_i^0 E\{\mb{u}\mb{u}^H\} \right) \\
	& = \mathrm{tr}(\mb{P}_i^0) - N^{-1} \mathcal{Q} \psi_0(\mathcal{Q}) \mathrm{tr}\left(\mb{P}_i^0 \right), \; i=1,\ldots,d. 
\end{split}
\ee

Finally, by substituting \eqref{score_vect_rep} and \eqref{proj_ope} in \eqref{semi_eff_score}, we get explicit expressions for the $d$ entries of the semiparametric efficient score vector $\bar{\mb{s}}_{\bs{\theta}_0}$ as:
\be
\label{eff_score_vect_rep}
\begin{split}
	[\bar{\mb{s}}_{\bs{\theta}_0}]_i &=_d \psi_0(\mathcal{Q}) \left( N^{-1}\mathcal{Q}  \mathrm{tr}\left(\mb{P}_i^0\right)  -2\sqrt{\mathcal{Q}}\mathrm{Re}\left[ \mb{u}^H \bs{\Sigma}_0^{-1/2}\bs{\mu}_i^0 \right]- \mathcal{Q}\mb{u}^H \mb{P}_i^0 \mb{u}\right) \\
	&=_d \psi_0(\mathcal{Q}) \left( N^{-1}\mathcal{Q}  \mathrm{tr}\left(\mb{P}_i^0\right) -\sqrt{\mathcal{Q}} \mb{u}^H \bs{\Sigma}_0^{-1/2}\bs{\mu}_i^0  - \sqrt{\mathcal{Q}} (\bs{\mu}_i^0)^H\bs{\Sigma}_0^{-1/2}\mb{u} -\mathcal{Q}\mb{u}^H \mb{P}_i^0 \mb{u}\right),
\end{split}
\ee
for $i = 1,\ldots,d$.

\subsubsection{Evaluation of the SFIM $\bar{\mb{I}}(\bs{\theta}_0|h_0)$}
As mentioned before, the SFIM for the estimation of $\bs{\theta}_0$ in the presence of the unknown, infinite-dimensional, nuisance parameter $h_0$ is given by $\bar{\mb{I}}(\bs{\theta}_0|h_0)\triangleq E_0\{\bar{\mb{s}}_{\bs{\theta}_0}\bar{\mb{s}}_{\bs{\theta}_0}^{H}\}$. In the sequel, a sketch of the calculation required to obtain an explicit expression for each entry of $\bar{\mb{I}}(\bs{\theta}_0|h_0)$ is reported.

Let us start by defining the vector $\mb{t} \triangleq \bs{\Sigma}_0^{-1/2}\mb{u}$ and then, substituting $\mb{t}$ in \eqref{eff_score_vect_rep}, we get:
\be
\begin{split}
	[\bar{\mb{s}}_{\bs{\theta}_0}]_i &=_d \psi_0(\mathcal{Q}) \left( N^{-1}\mathcal{Q}  \mathrm{tr}(\mb{P}_i^0) -\sqrt{\mathcal{Q}} \mb{t}^H \bs{\mu}_i^0 - \sqrt{\mathcal{Q}} (\bs{\mu}_i^0)^H\mb{t} -\mathcal{Q}\mb{t}^H \mb{\Sigma}_i^0 \mb{t}\right).
\end{split}
\ee
The next step consists in evaluating the products:
\be
\begin{split}
	[\bar{\mb{s}}_{\bs{\theta}_0}]_i&[\bar{\mb{s}}_{\bs{\theta}_0}]^*_j = \psi(\mathcal{Q})^2 \left[ N^{-2} \mathcal{Q}^2  \mathrm{tr}(\mb{P}_i^0)\mathrm{tr}(\mb{P}_j^0) - \right. \\
	&\left. - N^{-1}\mathcal{Q}^2 \left(   \mathrm{tr}(\mb{P}_i^0)\mb{t}^H \mb{\Sigma}_j^0 \mb{t} + \mathrm{tr}(\mb{P}_j^0)\mb{t}^H \mb{\Sigma}_i^0 \mb{t} \right)  \times \right. + \\ 
	& + \mathcal{Q} \mb{t}^H \bs{\mu}_i^0\mb{t}^H \bs{\mu}_j^0 + \mathcal{Q} \mb{t}^H \bs{\mu}_i^0(\bs{\mu}_j^0)^H\mb{t}+\\
	& + \mathcal{Q} (\bs{\mu}_i^0)^H\mb{t}\mb{t}^H\bs{\mu}_j^0 + \mathcal{Q} (\bs{\mu}_i^0)^H\mb{t}(\bs{\mu}_j^0)^H\mb{t} + \\
	& \left. + \mathcal{Q}^2 \mb{t}^H \mb{\Sigma}_i^0 \mb{t}\mb{t}^H \mb{\Sigma}_j^0 \mb{t} \right] \; i,j=1,\ldots,d.
\end{split}
\ee
Finally, by taking the expectation w.r.t. the true pdf $p_0(\mb{z})$ and by using the relations derived in (B.4)-(B.10) of \cite[Appendix B]{miss_sb}, it is easy to verify that each entry of the SFIM $\bar{\mb{I}}(\bs{\theta}_0|h_0)$ can be expressed as: 
\be
\label{SFIM_SB}
\begin{split}
	[\bar{\mb{I}}&(\bs{\theta}_0|h_0)]_{i,j} = \frac{ 2\bar{E}\{\mathcal{Q}\psi_0(\mathcal{Q})^2\}}{N} \mathrm{Re}[(\bs{\mu}_i^0)^H\bs{\Sigma}_0^{-1}\bs{\mu}_j^0]+ \\ 
	& + \frac{ \bar{E}\{\mathcal{Q}^2\psi_0(\mathcal{Q})^2\}}{N(N+1)} \left[ \mathrm{tr}(\bs{\Sigma}_0^{-1}\bs{\Sigma}_i^0\bs{\Sigma}_0^{-1}\bs{\Sigma}_j^0) - N^{-1}\mathrm{tr}(\bs{\Sigma}_0^{-1}\bs{\Sigma}_i^0)\mathrm{tr}(\bs{\Sigma}_0^{-1}\bs{\Sigma}_j^0) \right],
\end{split}
\ee
for $i,j=1,\ldots,d$ and where $\bar{E}\{\cdot\}$ is defined in \eqref{bar_E}.

\subsubsection{A compact expression for $\bar{\mb{I}}(\bs{\theta}_0|h_0)$}
\label{compact_ex}
Using the well-known properties of the Kronecker product $\otimes$ and of the standard vectorization operator $\mathrm{vec}$ (see e.g. \cite{Magnus1, Magnus2}), it is possible to rewrite the SFIM in \eqref{SFIM_SB} in a more compact and easy-to-use form. This expression will represent the SSB formula for a single CES-distributed snapshot.

Let us define two Jacobian matrices of the mean vector $\bs{\mu}(\bs{\theta})$ and of the scatter matrix $\bs{\Sigma}(\bs{\theta})$ as $\mb{N}_0 = \nabla_{\bs{\theta}}^T\bs{\mu}(\bs{\theta}_0) \in \mathbb{C}^{N \times d}$ and $\mb{V}_0 = \nabla_{\bs{\theta}}^T\mathrm{vec}(\bs{\Sigma}(\bs{\theta}_0)) \in \mathbb{C}^{N^2 \times d}$, respectively. Note that both $\mb{N}_0$ and $\mb{V}_0$ are evaluated at the true parameter vector $\bs{\theta}_0$. Then, the $\bar{\mb{I}}(\bs{\theta}_0|h_0)$ can be written in a compact Gramian form as:
\be
\label{Gram_SFIM}
\begin{split}
	\bar{\mb{I}}(\bs{\theta}_0|h_0)&=\frac{ 2\bar{E}\{\mathcal{Q}\psi_0(\mathcal{Q})^2\}}{N} \mathrm{Re}[(\bs{\Sigma}_0^{-1/2}\mb{N}_0)^H(\bs{\Sigma}_0^{-1/2}\mb{N}_0)] +\\
	& + \frac{ \bar{E}\{\mathcal{Q}^2\psi_0(\mathcal{Q})^2\}}{N(N+1)} (\mb{T}^{1/2}\mb{V}_0)^H(\mb{T}^{1/2}\mb{V}_0) \\
	& = \frac{ 2\bar{E}\{\mathcal{Q}\psi_0(\mathcal{Q})^2\}}{N} \mathrm{Re}[\mb{N}_0^H\bs{\Sigma}_0^{-1}\mb{N}_0] + \frac{ \bar{E}\{\mathcal{Q}^2\psi_0(\mathcal{Q})^2\}}{N(N+1)} \mb{V}_0^H\mb{T}\mb{V}_0,
\end{split}
\ee
where $\bar{E}\{\cdot\}$ is defined in \eqref{bar_E} and the matrices $\mb{T}^{1/2}$ and $\Pi^{\perp}_{\cvec{\mb{I}_N}}$ are:
\be
\label{sqrt_T}
\mb{T}^{1/2} = \Pi^{\perp}_{\cvec{\mb{I}_N}}(\kronsigmatinvmT),
\ee
\be
\Pi^{\perp}_{\cvec{\mb{I}_N}}=\mb{I}_{N^2} - N^{-1}\mathrm{vec}(\mb{I}_N)\mathrm{vec}(\mb{I}_N)^T.
\ee
As the notation suggests, matrix $\Pi^{\perp}_{\cvec{\mb{I}_N}}$ is the orthogonal projection matrix on the orthogonal complement of $\mathrm{span}(\mathrm{vec}(\mb{I}_N))$. Then, by exploiting the property of $\otimes$ and the fact that an orthogonal projection matrix is idempotent, we have that: 
\be
\label{T_mat}
\mb{T} \triangleq \bs{\Sigma}_0^{-T} \otimes \bs{\Sigma}_0^{-1} - N^{-1} \mathrm{vec}(\bs{\Sigma}_0^{-1})\mathrm{vec}(\bs{\Sigma}_0^{-1})^H.
\ee

\textit{Remark}: It is worth noticing that the compact expression of $\bar{\mb{I}}(\bs{\theta}_0|h_0)$ obtained in \eqref{Gram_SFIM} encompasses as special cases the expressions of the SFIM for the scatter matrix estimation derived in \cite[eq. 56]{For_SCRB}. To clarify this point, let us consider the scatter matrix estimation problem under the assumption of a perfectly known mean vector. Since in the RES case the scatter matrix $\bs{\Sigma}_0$ is a real (symmetric) matrix, then the unknown parameter vector can be recast as $\bs{\theta}_0 = \mathrm{vecs}(\bs{\Sigma}_0)$, where the $\mathrm{vecs}$ operator maps the symmetric $N \times N$ matrix $\bs{\Sigma}_0$ to an $N(N+1)/2$-dimensional vector containing the elements of the lower triangular sub-matrix of $\bs{\Sigma}_0$. This definition of $\bs{\theta}_0$ implies that the Jacobian matrix of the mean vector $\mb{N}_0$ is nil while the Jacobian matrix of the scatter matrix is given by $\mb{V}_0 = \nabla_{\mathrm{vecs}(\bs{\Sigma}_0)}^T\mathrm{vec}(\bs{\Sigma}_0) = \mb{D}_N$, where $\mb{D}_N$ is the so-colled \textit{duplication matrix} and the last equality follows from \cite[Lemma 3.8]{Magnus2}. Finally, by substituting the derived expressions for the two Jacobian matrices in \eqref{Gram_SFIM}, we immediately obtain the expression of the SFIM for the (real) scatter matrix estimation problem already derived in \cite[eq. 56]{For_SCRB}.

\subsection{The Multiple Snapshot Case}
\label{multi_snap}
In this subsection, we provide two extensions of the single-snapshot SSB formula derived in Sec. \ref{single_snap} to two multi-snapshot scenarios. Before starting with the derivation, a comment is in order. The general multiple-snapshot scenario considered in array processing applications is characterized by the availability of $L$ independent, CES-distributed, data vectors $\mb{z}_l \sim CES_{N}(\mb{z}_l;\bs{\mu}_l(\bs{\theta}_0), \bs{\Sigma}(\bs{\theta}_0),h_0)$ sharing the same scatter matrix but with a possibly different mean vector from snapshot to snapshot\footnote{To clarify this point, one could thing to the \virg{deterministic signal model} commonly used in array processing where the mean vector of the observations is modelled as $\bs{\mu}_l(\bs{\theta}_0) = \alpha_l\mb{a}(\bs{\theta}_0)$ where $\mb{a}(\bs{\theta}_0)$ is the steering vector and $\alpha_l$ is a deterministic (generally unknown) complex scalar that changes from snapshot to snapshot.}. Due to the possible variation of the mean vectors, the available data $\{\mb{z}_l\}_{l=1}^L$ are not identically distributed. In particular, the semiparametric group structure of the set
$\mathcal{P}_{\bs{\theta},h}$ in \eqref{CES_semi_group_model} no longer holds, since the affine transformations in \eqref{affine_tran} will depends on $l$. The extension of the classical semiparametric theory to the non-i.i.d. (\textit{independent} and \textit{identical distributed}) case is a well established topic (see e.g. \cite{non_iid} for a summary of main works in this filed or the seminal paper \cite{Hallin_Werker}), but falls outside the scope of this paper. Consequently, we left the extension of the SSB formula in \eqref{Gram_SFIM} to the general, non-i.i.d. multiple snapshot case for future work, while here we focus our attention on two less general models which, however, are still relevant in practice.

\subsubsection{SSB formula for the Elliptical Vector (EV) Model}
The so-called Elliptical Vector model has been already used in \cite{RichPhD}, \cite{Rich_letter}, \cite{Rich} and in \cite{Bess}. Specifically, in \cite{Rich} the EV model has been exploited to derive the \textit{misspecified} SB formula under the mismatched Gaussian assumption. The basic idea behind the EV model is to consider as snapshots the $L$ sub-vectors of an $LN$-dimensional, CES-distributed, random vector. More formally, suppose to have an $LN$-dimensional CES-distributed vector $\mathbb{C}^{LN} \ni \mb{z}  \triangleq [\mb{z}_1^T,\ldots,\mb{z}_L^T]^T\sim CES_{LN}(\mb{z};\bs{\mu}_0, \bs{\Sigma}_0,h_0)$ whose mean vector and scatter matrix are defined as:  
\be
\label{tilde_mu}
\bs{\mu}_0 \triangleq [\bs{\mu}_1(\bs{\theta}_0)^T,\ldots,\bs{\mu}_L(\bs{\theta}_0)^T]^T \equiv [\bs{\mu}_{1,0}^T,\ldots,\bs{\mu}_{L,0}^T]^T \in \mathbb{C}^{LN},
\ee
\be
\label{tilde_sigma}
\bs{\Sigma}_0 \triangleq \mb{I}_L \otimes \bs{\Omega}(\bs{\theta}_0) \equiv \mb{I}_L \otimes \bs{\Omega}_0 \in \mathbb{C}^{LN \times LN}.
\ee
Under these assumptions on $\mb{z}$, we can use \cite[Lemma 3.5]{CES_stat} and \cite[Theo. 2]{Esa} to derive some useful properties of the sub-vectors $\{\mb{z}_l\}_{l=1}^L$. Specifically, for each $l$, $\mb{z}_l \sim CES_{N}(\mb{z}_l;\bs{\mu}_{l,0}, \bs{\Omega}_0,\tilde{h}_0)$ is an $N$-dimensional CES-distributed random vector with mean vector $\bs{\mu}_{l,0}$, scatter matrix $\bs{\Omega}_0$ and \virg{marginal} density generator $\tilde{h}_0$ that is related to $h_0$ by the integral equation given in \cite[eq. 3.89]{RichPhD}. It is important to note that, even if, in general, the functional form of $h_0$ is different from the one of its \virg{marginal} counterpart $\tilde{h}_0$, the vector $\mb{z}$ and all its sub-vectors $\{\mb{z}_l\}_{l=1}^L$ share the same \textit{characteristic generator} \cite[Theo. 2]{Esa}. From \cite[Lemma 3.5]{CES_stat}, $\mb{z}_l$ admits the following stochastic representation: $\mb{z}_l - \bs{\mu}_{l,0} =_d \sqrt{\mathcal{Q}_l} \bs{\Omega}_0^{1/2} \mb{u}_l$, $\forall l=1,\ldots,L$, where $\mathbf{u}_l \sim U(\mathbb{C}S^{N})$ is independent of $\mathcal{Q}_l$. Furthermore, $\mathcal{Q}_l =_d \beta \mathcal{Q}$ where $\beta \sim \mathrm{Beta}(N,N(L-1))$ is a Beta-distributed random variable, independent of $\mathcal{Q}$ that is the second-order modular variate of $\mb{z}$. The derivation of the SSB formula for the EV model can be be easily obtained by substituting the expressions of $\bs{\mu}_0$ and $\bs{\Sigma}_0$, given in \eqref{tilde_mu} and \eqref{tilde_sigma}, in the SSB formula already derived in \eqref{SFIM_SB}. Finally, by using the properties of the Kronecker product, we get:
\be
\label{SSB_EV}
\begin{split}
	[\bar{\mb{I}}_L(\bs{\theta}_0|g_0)]_{i,j} &= \frac{ 2\bar{E}\{\mathcal{Q}\psi_0(\mathcal{Q})^2\}}{LN} \sum_{l=1}^{L}\mathrm{Re}[(\bs{\mu}_{i,l}^0)^H\bs{\Sigma}_0^{-1}\bs{\mu}_{j,l}^0]+  \\ 
	& + \frac{ \bar{E}\{\mathcal{Q}^2\psi_0(\mathcal{Q})^2\}}{N(LN+1)} \left[ \mathrm{tr}(\bs{\Omega}_0^{-1}\bs{\Omega}_i^0\bs{\Omega}_0^{-1}\bs{\Omega}_j^0) -N^{-1}\mathrm{tr}(\bs{\Omega}_0^{-1}\bs{\Omega}_i^0)\mathrm{tr}(\bs{\Omega}_0^{-1}\bs{\Omega}_j^0) \right],
\end{split}
\ee
for $i,j=1,\ldots,d$ and where $\bar{E}\{\cdot\}$ is defined in \eqref{bar_E}. Clearly, this expression of the SFIM for the SV model can be rewritten in a compact Gramian form following the same procedure used in Sec. \ref{compact_ex}.

\subsubsection{Semiparametric Bangs formula}
\label{sem_bangs}
Assume to have a set of $L$ i.i.d. CES-distributed random vectors $\{\mb{z}_l\}_{l=1}^L$ sampled from $CES_{N}(\mb{z}_l;\bs{\mu}_l, \bs{\Sigma}(\bs{\theta}_0),h_0)$, where the mean vector is assumed to be constant with respect to $\bs{\theta}$. Since the data are i.i.d. random vectors, the multiple-snapshot extension of the SSB formula in \eqref{Gram_SFIM} is trivial. Let us define the multi-snapshot SFIM as  $\bar{\mb{I}}_L(\bs{\theta}_0|g_0) \triangleq E_0\{\bar{\mb{s}}_{\bs{\theta}_0}(\{\mb{z}_l\}_{l=1}^L)\bar{\mb{s}}_{\bs{\theta}_0}(\{\mb{z}_l\}_{l=1}^L)^H\}$, then, from \eqref{Gram_SFIM}, we have:
\be
\label{Multi_Gram_SFIM}
\bar{\mb{I}}_L(\bs{\theta}_0|h_0) = L\frac{ \bar{E}\{\mathcal{Q}^2\psi_0(\mathcal{Q})^2\}}{N(N+1)} \mb{V}_0^H\mb{T}\mb{V}_0,
\ee
where the function $\psi_0$ has already been defined in \eqref{psi}, and matrices $\mb{V}_0$ and $\mb{T}$ have been defined in Sec. \ref{compact_ex}.

In the next section, we show how to apply the SSB formula in \eqref{Multi_Gram_SFIM} to a well-know problem in array processing.

\section{The Semiparametric Stochastic CRB for array processing}

This section is dedicated to the derivation of the semiparametric version of the well-known Stochastic CRB for DOA estimation problems under random signal models \cite{Stoica_CRB,Stoica_CRB_2,Ottersten,Weiss,Renaux}.

Assume to have an array of $N$ sensors and $K$ narrowband sources impinging on the array and characterized by $\{\nu_1,\ldots,\nu_K\}$ direction parameters. Let us assume to collect $L$ i.i.d. and CES-distributed data snapshots $\{\mb{z}_l\}_{l=1}^L$, such that $\mb{z}_l \sim CES_{N}(\mb{z};\mb{0}, \bs{\Sigma}(\bs{\nu},\bs{\Gamma},\sigma^2), h_0), \; \forall l$ where the density generator $h_0 \in \bar{\mathcal{G}}$, that is constrained as in \eqref{const}, is left unspecified, and \cite{Esa_DOA}:
\be
\label{Cov_mat_DOA}
\bs{\Sigma} \equiv \bs{\Sigma}(\bs{\nu},\bs{\Gamma},\sigma^2) = \mb{A}(\bs{\nu})\bs{\Gamma}\mb{A}(\bs{\nu})^H + \sigma^2\mb{I}_N.
\ee
where:
\begin{itemize}
	\item $\mb{A}(\bs{\nu}) \triangleq [\mb{a}(\nu_1) \cdots \mb{a}(\nu_K)]$ is the steering matrix with $\bs{\nu} \triangleq (\nu_1,\ldots,\nu_K)^T$ and $\mb{a}(\nu_k)$ is the array steering vector for the $k$-th source,
	\item $\bs{\Gamma}$ is the source covariance matrix,
	\item $\sigma^2$ is the noise power.
\end{itemize} 

For the subsequent derivation, it is useful to introduce the vector $\bs{\zeta}$ as the $N^2$-dimensional real vector such that:
\be
\bs{\zeta} \triangleq \tonde{\mathrm{diag}(\bs{\Gamma})^T,\mathrm{vec}_l(\mathrm{Re}(\bs{\Gamma}))^T,\mathrm{vec}_l(\mathrm{Im}(\bs{\Gamma}))^T}^T,
\ee
 where the operator $\mathrm{vec}_l(\cdot)$ is defined as in \eqref{def_theta}.

Let us now collect in the $(K+N^2+1)$-dimensional vector
\be
\bs{\theta} \triangleq [\bs{\nu}^T,\bs{\zeta}^T,\sigma^2]^T
\ee
all the finite-dimensional unknown parameters. Note that, in general, we are interested only in the estimation of $\bs{\nu}$, while the signal covariance matrix $\bs{\Gamma}$ (or, equivalently $\bs{\zeta}$) and the noise power $\sigma$ have to be considered as nuisance terms. Following the notation introduced in the previous sections, the \textit{true} parameter vector will be indicated as $\bs{\theta}_0 = [\bs{\nu}^T_0,\bs{\zeta}^T_0,\sigma_0^2]^T$. Similarly, the true signal covariance matrix will be indicated as $\bs{\Gamma}_0$.

The SFIM for the estimation of $\bs{\theta}_0$ can be directly obtained by applying the semiparametric Bangs formula given in \eqref{Multi_Gram_SFIM} as:
\be
\label{sto_8}
\begin{split}
	\bar{\mb{I}}_L(\bs{\theta}_0|h_0)
	= L\frac{ \bar{E}\{\mathcal{Q}^2\psi_0(\mathcal{Q})^2\}}{N(N+1)} \quadre{\mb{T}^{1/2}\nabla^T_{\bs{\theta}}\mathrm{vec}(\bs{\Sigma}(\bs{\theta}_0))}^H \quadre{\mb{T}^{1/2}\nabla^T_{\bs{\theta}}\mathrm{vec}(\bs{\Sigma}(\bs{\theta}_0))},
\end{split}
\ee 
where $\mb{T}^{1/2}$ has been introduced in \eqref{sqrt_T} as $\mb{T}^{1/2} = \Pi^{\perp}_{\cvec{\mb{I}_N}}(\kronsigmatinvmT)$.
It is immediate to verify that \eqref{sto_8} is the semiparametric counterpart of \cite[eq. (8)]{Stoica_CRB}.

Similarly to \cite[eq. (10)]{Stoica_CRB}, let us define the matrices $\mb{G}_s$ and $\bs{\Delta}_s$ as:
\be
\label{sto_10}
\begin{split}
	\mb{T}^{1/2}&\nabla^T_{\bs{\theta}}\mathrm{vec}(\bs{\Sigma}(\bs{\theta}_0))\\ &=\mb{T}^{1/2}\quadre{\nabla^T_{\bs{\nu}}\mathrm{vec}(\bs{\Sigma}(\bs{\theta}_0)), \nabla^T_{\bs{\zeta}}\mathrm{vec}(\bs{\Sigma}(\bs{\theta}_0)),\parder{\mathrm{vec}(\bs{\Sigma}(\bs{\theta}_0))}{\sigma} }\\
	& \triangleq \quadre{\mb{G}_s,\bs{\Delta}_s} = \quadre{\Pi^{\perp}_{\cvec{\mb{I}_N}}\mb{G},\Pi^{\perp}_{\cvec{\mb{I}_N}}\bs{\Delta}},
\end{split}
\ee
where the matrices $\mb{G}$ and $\bs{\Delta}$ are implicitly defined by the second equality in \eqref{sto_10} and are the same of the ones in \cite[eq. (10)]{Stoica_CRB}.

By substituting \eqref{sto_10} in \eqref{sto_8}, we get that the SFIM in \eqref{sto_8} can be expressed in the following block-matrix form:
\be
\label{sto_11}
\bar{\mb{I}}_L(\bs{\theta}_0|h_0) = L\frac{ \bar{E}\{\mathcal{Q}^2\psi_0(\mathcal{Q})^2\}}{N(N+1)}\left( 
\begin{array}{cc}
	\mb{G}^H_s\mb{G}_s & \mb{G}_s^H\bs{\Delta}_s\\
	\bs{\Delta}_s^H\mb{G}_s & \bs{\Delta}_s^H\bs{\Delta}_s
\end{array}
\right) .
\ee

Since, as said before, we are interested only in the estimation of the direction parameter vector $\bs{\nu}_0$, the relevant expression of the SCRB is given by the top-left $K \times K$ submatrix of the inverse of \eqref{sto_11}. By using the Woodbury identity \cite[eq. (157)]{Matrix_Cook}, this submatrix, that represent the Semiparametric Stochastic CRB (SSCRB) can be obtained as: 
\be
\label{sto_12}
\begin{split}
\mathrm{SSCRB}(\bs{\nu}_0|\bs{\zeta}_0,\sigma_0^2, h_0) &= \frac{N(N+1)}{L\bar{E}\{\mathcal{Q}^2\psi_0(\mathcal{Q})^2\}} \quadre{\mb{G}^H_s\mb{G}_s - \mb{G}^H_s\bs{\Delta}_s\tonde{\bs{\Delta}_s^H\bs{\Delta}_s}^{-1}\bs{\Delta}_s^H\mb{G}_s}^{-1}\\
& = \frac{N(N+1)}{L\bar{E}\{\mathcal{Q}^2\psi_0(\mathcal{Q})^2\}} \quadre{\mb{G}^H_s \Pi^{\perp}_{\bs{\Delta}_s}\mb{G}_s}^{-1},
\end{split} 
\ee
that represents the semiparametric counterpart of \cite[eq. (12)]{Stoica_CRB}. It is possible to show (see the proof in the Appendix) that:
\be
\label{SSCRB}
\mathrm{SSCRB}(\bs{\nu}_0|\bs{\zeta}_0,\sigma_0^2, h_0) = \frac{N(N+1)\sigma_0^2}{2L\bar{E}\{\mathcal{Q}^2\psi_0(\mathcal{Q})^2\}}\quadre{\mathrm{Re}\tonde{\mb{D}_0^H \Pi^{\perp}_{\mb{A}_0} \mb{D}_0}\odot \tonde{\bs{\Gamma}_0\mb{A}_0^H\mb{\Sigma}_0^{-1}\mb{A}_0\mb{\Gamma}_0}^T}^{-1},
\ee
where $\odot$ is the Hadamard product, $\bar{E}\{\cdot\}$ is defined as in \eqref{bar_E} and $\mb{D}_0 \triangleq \quadre{\mb{d}_{0,1},\cdots,\mb{d}_{0,K}}$ where $\mb{d}_{0,k}$ is
\be
\label{der_steer}
\mb{d}_{0,k} \triangleq \left. \frac{d\mb{a}(\nu_k)}{d\nu_k} \right|_{\nu_k = \nu_{0,k}}.
\ee
To conclude, we note that eq. \eqref{der_steer} can be easily extended to the case in which, instead of the diagonal matrix $\sigma_0^2\mb{I}_N$, the noise covariance matrix in \eqref{Cov_mat_DOA} is non-diagonal by following the procedures detailed in \cite{Non_diag_Noise}.

\section{Numerical results}
The aim of this section is to provide some numerical examples that can help to clarify the practical usefulness of the theoretical findings. In subsection \ref{num_t_CCSCRB}, we show how to calculate the constrained CSCRB derived in Sec. \ref{CRB_sec} for a set of complex, $t$-distributed random vectors and we investigate the efficiency of two popular (constrained) scatter matrix estimators, i.e. the Sample Covariance Matrix (SCM) and the Tyler's estimator. Secondly, in subsection \ref{SSCRB_example}, an example regarding the use of the SSCRB in \eqref{SSCRB} as a bound for the MSE of the adaptive MUSIC DOA estimator in $t$-distributed data is discussed.
\subsection{CCSCRB for $t$-distributed data}
\label{num_t_CCSCRB}
The pdf related to the complex $t$-distribution can be obtained from the real $t$-distribution by applying the equality chain in \eqref{RES_CES}. Specifically, the relevant density generator $h_0$ can be obtained from the one given in eq. (75) in \cite{For_SCRB} through a change of variables $N \rightarrow 2N$, $\lambda \rightarrow \lambda/2$ as: 
\be
\label{dg_t_dist}
h_0(t) =(\pi^{N}\Gamma({\lambda} ))^{-1}\Gamma(\lambda+N)(\lambda/\eta) ^{\lambda}(\lambda/\eta + t)^{-(\lambda+N)} 
% \frac{\Gamma(\lambda+N)}{\pi^{N}\Gamma({\lambda} )}\left( \frac{\lambda}{\eta}\right) ^{\lambda}\left( \frac{\lambda}{\eta} + t \right)^{-(\lambda+N)} 
\ee
and then $\psi_0(t) = -(\lambda + N) (\lambda/\eta + t) ^ {-1}.$ From \eqref{CSS_Q_pdf}, we have that:
\be
p_{\mathcal{Q}}(q) = \frac{\Gamma(\lambda+N)}{\Gamma(N)\Gamma({\lambda} )}\left( \frac{\lambda}{\eta}\right)^{\lambda}q^{N-1}\left( \frac{\lambda}{\eta} + q \right)^{-(\lambda+N)}.
\ee
Using the integral in \cite[pp. 315, n. 3.194 (3)]{Integrals}, we get:
\be
\label{exp_1}
E\{\mathcal{Q}\psi_0(\mathcal{Q})^2\}  =  \frac{\eta N (\lambda + N)}{N +\lambda + 1},
\ee
\be
\label{exp_2}
E\{\mathcal{Q}^2\psi_0(\mathcal{Q})^2\} =  \frac{ N (N+1) (\lambda + N)}{(N +\lambda +1)}.
\ee
Finally, by inserting \eqref{exp_1} and \eqref{exp_2} in \eqref{FIM_mu} and \eqref{inv_SFIM_com_Sigma}, we obtain closed form expressions for the matrices $\mb{C}_0(\bar{\mb{s}}_{\bs{\mu}_0})$ and $\mb{C}_0(\bar{\mb{s}}_{\mathrm{vec}(\bs{\Sigma}_0)})$ and consequently the CCSCRB in \eqref{SCRB_gun_com}.

\begin{figure}[H]
	\centering
	\includegraphics[height=6cm]{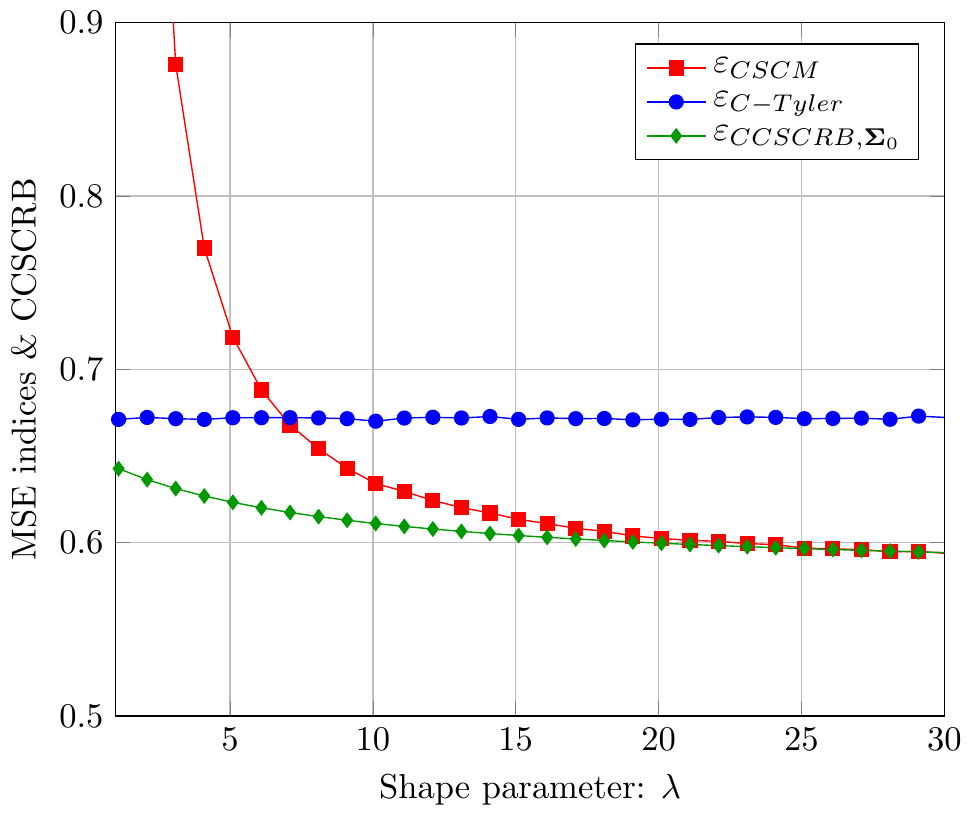}
	\caption{MSE indices for the CSCM and C-Tyler's estimators and the related CCSCRB as functions of the shape parameter $\lambda$ for complex \textit{t}-distributed data ($L=3N$).}
	\label{fig:Fig1}
\end{figure}

In Fig. \ref{fig:Fig1}, the performance of the constrained SCM (CSCM) estimator and the constrained Tyler's (C-Tyler) estimator are compared against the CCSCRB. The explicit expressions of these two estimators can be obtained from those provided in \cite{For_SCRB} for real data by replacing the transpose with the Hermitian operator. The simulation parameters are:
\begin{itemize}
	\item $\bs{\Sigma}_0$ is a Toeplitz Hermitian matrix whose first column is given by $[1,\rho, \ldots,\rho^{N-1}]^T$, where $\rho = 0.8e^{j2\pi/5}$ and $N=8$.
	\item The data power is $\sigma_X^2 = E\{\mathcal{Q}\}/N = 4$.
	\item The data is assumed to be zero mean, i.e. $\bs{\mu}_0=\mb{0}_N$.
	\item The number of the available i.i.d. data vectors is $L=3N=24$. Since we assume to have $L$ i.i.d. data vectors, the CCSCRB in \eqref{SCRB_gun_com} has to be divided by $L$.
	\item The number of independent Monte Carlo runs is $10^6$.
\end{itemize}
As MSE indices and bound, in Fig. \ref{fig:Fig1} we plot:
\be
\varepsilon_\alpha \triangleq \norm{E\{(\mathrm{vec}(	\hat{\bs{\Sigma}}_\alpha)-\mathrm{vec}(\bs{\Sigma}_0))(\mathrm{vec}(	\hat{\bs{\Sigma}}_\alpha)-\mathrm{vec}(\bs{\Sigma}_0))^H\}}_F,
\ee
where $\alpha = \{CSCM,C-Tyler\}$ and 
\be
\varepsilon_{CCSCRB,\bs{\Sigma}_0} \triangleq \norm{[\mathrm{CCSCRB}(\bs{\phi}_0,h_0)]_{\bs{\Sigma}_0}}_F.
\ee
In Fig. \ref{fig:Fig1} we compare the MSE of the CSCM and C-Tyler's estimators with the CCSCRB as function of the shape parameter $\lambda$. When $\lambda \rightarrow \infty$, i.e. when the data tends to be Gaussian distributed, the CSCM tends to the CSCRB. Fig. \ref{fig:Fig1} shows that the C-Tyler's estimator is not an efficient estimator w. r. t. the CCSCRB, even if its performance is higher than that of the SCM for highly non-Gaussian data (i.e. small $\lambda$). Moreover, since the C-Tyler's estimator is a robust estimator, its MES is invariant w.r.t. the shape parameter, as expected. 

\subsection{Semiparametric Stochastic CRB and MUSIC algorithm}
\label{SSCRB_example}
In this subsection, we show how to apply the SSCRB given in \eqref{SSCRB} in a simple but representative problem in array processing. We assume to have a uniformly linear array (ULA) of N omnidirectional sensors and a single ($K=1$) narrowband source impinging on the array with spatial frequency $\nu_0$.\footnote{For the ULA configuration, the spatial frequency is defined as $\nu = d/\lambda\sin(\theta)$ where $d$ is the spacing between the sensor, $\lambda$ is the wavelength of the transmitted signal and $\theta$ is the conic angle of the source.} Note that, for a ULA, the steering vector can be expressed as $\mb{a}(\nu_0)=[1,e^{j2\pi\nu_0},\ldots,e^{j2\pi(N-1)\nu_0}]^T$. We suppose to collect $L$, i.i.d. $t$-distributed data snapshots $\{\mb{z}_l\}_{l=1}^L$ whose scatter matrix is of the form given in \eqref{Cov_mat_DOA}:
\be
\label{Cov_single_source}
\bs{\Sigma}(\nu_0,\gamma_0^2,\sigma_0^2) = \gamma_0^2\mb{a}(\nu_0)\mb{a}(\nu_0)^H + \sigma_0^2\mb{I}_N,
\ee
where $\gamma_0^2$ is the (unknown) power of the single source impinging on the array while $\sigma_0^2$ is the (unknown) power of the white noise component. It is worth highlighting that, as discussed in Sec. \ref{SSB_sec}, we assume that the density generator of the $t$-distribution, given in \eqref{dg_t_dist}, satisfies the constraint in \eqref{const}, so that the scatter matrix in \eqref{Cov_single_source} is the covariance matrix of $\mb{z}_l,\; \forall l$. It is immediate to verify that the constraint is satisfied by choosing $\eta = \lambda/(\lambda-1)$.

\begin{figure}[H]
	\centering
	\includegraphics[height=6cm]{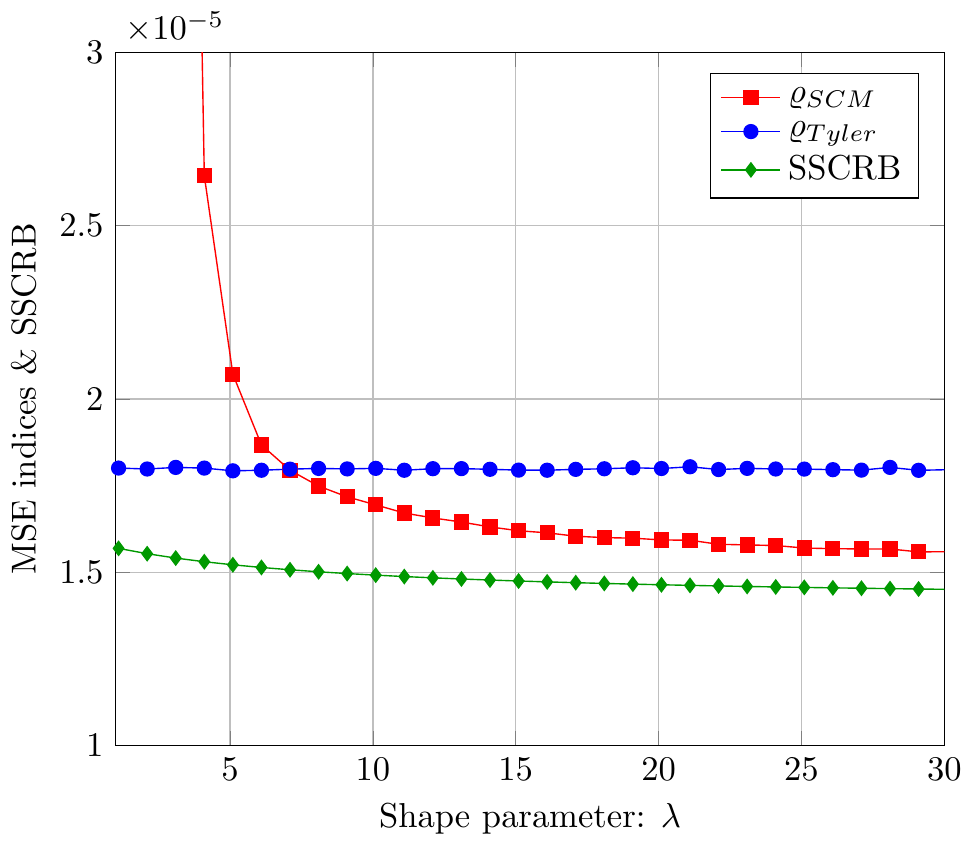}
	\caption{MSE indices for the MUSIC-SCM and MUSIC-Tyler spatial frequency estimators and the related SSCRB as functions of the shape parameter $\lambda$ for complex \textit{t}-distributed data ($L=3N$).}
	\label{fig:Fig2}
\end{figure}

The parameter of interest that has to be estimated is the source spatial frequency $\bs{\nu}$, while $\gamma_0$ and $\sigma_0^2$ represent two (finite-dimensional) nuisance parameters. To estimate $\nu$, we adopt the MUSIC algorithm (see e.g. \cite{MUSIC}):
\be
\label{Mus_al}
\hat{\nu} = \underset{\nu}{\mathrm{argmax}} \graffe{\quadre{\sum\nolimits_{n=K+1}^N |\mb{a}(\nu)^H\hat{\mb{v}}_n|^2}^{-1}},
\ee  
where $\mb{a}(\nu)$ is the steering vector and $\{\hat{\mb{v}}_n\}_{n=K+1}^N$ are the $N-K$ eigenvectors corresponding to the $N-K$ smallest eigenvalues of the estimated data covariance matrix $\hat{\bs{\Sigma}}$. In the following, we assess the efficiency w.r.t. the SSCRB of the MUSIC algorithm in \eqref{Mus_al} when the unknown covariance matrix $\bs{\Sigma}$ in \eqref{Cov_single_source} is estimated by means of the SCM or Tyler's estimators. Note that none of these two estimators rely on the knowledge of the density generator. As MSE indices and bound we use:
\be
\varrho_\alpha \triangleq E\{(\hat{\nu}_\alpha - \nu_0)^2\},
\ee
where $\alpha = \{SCM,Tyler\}$, while the bound is $\mathrm{SSCRB}(\nu_0|\delta_0,\sigma_0^2, h_0)$ obtained by specializing the general expression in \eqref{SSCRB} for the particular case at hand. Note that, for the $t$-distribution, the expectation operator $\bar{E}\{\mathcal{Q}^2\psi_0(\mathcal{Q})^2\}$ in \eqref{SSCRB} is equal to the one already evaluated in \eqref{exp_2}.
The simulation parameters are:
\begin{itemize}
	\item Spatial frequency $\nu_0 = 0.3$,
	\item The noise power $\sigma_0^2 = 1$ while the signal power $\gamma_0$ is chosen in order to have a Signal-to-Noise ration of 0 dB.
	\item $N=8$ and $L=3N=24$.
	\item The number of Monte Carlo runs in $10^6$.
\end{itemize}
In Fig. \ref{fig:Fig2}, we compare the MSE of two version of the MUSIC estimator with the SSCRB, as function of the shape parameter $\lambda$. Similar to the results in Fig. \ref{fig:Fig1}, the MUSIC-Tyler estimator achieves better performance when the data snapshot are highly non-Gaussian (small $\lambda$), and its MSE is invariant w.r.t. the shape parameter. On the other hand, the non-robust MUSIC-SCM estimator overtakes the MUSIC-Tyler estimator when the data tend to be Gaussian (large $\lambda$). However, neither the MUSIC-Tyler nor MUSIC-SCM estimators are efficient estimators w.r.t. the SSCRB.

\section{Conclusion}
In this paper, the Semiparametric CRB (SCRB) and related results, recently obtained for the RES model \cite{For_SCRB}, have been extended to the CES distributions. Specifically, we derived the SCRB for the (constrained) estimation of the complex mean vector and complex scatter matrix of a CES-distributed random vector. The proposed complex CSCRB is a lower bound on the estimation accuracy of any estimator of $\bs{\mu}$ and $\bs{\Sigma}$ when the density generator of the underlying CES distribution is unknown. Secondly, the Semiparametric Slepian-Bangs (SSB) formula for the estimation of a parameter vector $\bs{\theta}$ parametrizing the complex mean vector $\bs{\mu}(\bs{\theta})$ and the complex scatter matrix $\bs{\Sigma}(\bs{\theta})$ has been derived for CES-distributed data. Moreover, the proposed SSB formula has been exploited to obtain the semiparametric version of the Stochastic CRB for DOA estimation under random signal model assumption. Finally, some numerical results have been described with the aim of clarifying the practical usefulness of our theoretical findings. A lot of potential applications of the SCRB and the SSB formula to many Signal Processing problems still remain to be investigated. Along with its practical exploitation, the Semiparametric CRB  poses a series of theoretical questions as well including, in particular, the existence of an optimal trade-off between the semiparametric efficiency and the robustness of the estimator. 

\section*{Acknowledgment}
The work of Stefano Fortunati has been partially supported by the Air Force Office of Scientific Research under award number FA9550-17-1-0065.

\appendix[Derivation of the Semiparametric Stochastic CRB in eq. (62)]

This document is entirely dedicated to the calculation of $\mathrm{SSCRB}(\bs{\nu}_0|\bs{\zeta}_0,\sigma_0^2, h_0)$ given in (62) of our paper. To this end, whenever it is possible, we will rely on similar calculation already derived in \cite{Stoica_CRB} and then we will try to remain as close as possible to the notation used there. 

The crucial step for the calculation of $\mathrm{SSCRB}(\bs{\nu}_0|\bs{\zeta}_0,\sigma_0^2, h_0)$ is the evaluation of $\Pi^{\perp}_{\bs{\Delta}_s}$. According to \cite[eq. (13)]{Stoica_CRB}, we note that:
\be
\label{sto_13}
\bs{\Delta}_s = \Pi^{\perp}_{\cvec{\mb{I}_N}} \bs{\Delta} = \quadre{\Pi^{\perp}_{\cvec{\mb{I}_N}}\tilde{\mb{V}},\Pi^{\perp}_{\cvec{\mb{I}_N}}\mb{u}},
\ee
where $\tilde{\mb{V}}$ is defined as \cite[eq. (19)]{Stoica_CRB}:
\be
\label{sto_19}
\tilde{\mb{V}} \triangleq \mb{V}\mb{J} = (\bs{\Sigma}_0^{-T/2}\mb{A}^* \otimes \bs{\Sigma}_0^{-1/2}\mb{A})\mb{J},
\ee
where $\mb{J}$ is a non-singular matrix\footnote{More specifically, as discussed in \cite{Stoica_CRB}, $\mb{J}$ represents a change of basis. However, since its explicit form is immaterial for the subsequent derivation, we will not provide other details about it.} and $\mb{u}$ is given by \cite[eq. (22)]{Stoica_CRB}:
\be
\label{sto_22}
\mb{u} \triangleq \mb{T}^{1/2} \parder{\mathrm{vec}(\bs{\Sigma}(\bs{\theta}_0))}{\sigma} = \Pi^{\perp}_{\cvec{\mathbf{I}_N}}\cvec{\bs{\Sigma}_0^{-1}}.
\ee

Since the matrices $\bs{\Delta}_s$ and $\quadre{\Pi^{\perp}_{\cvec{\mb{I}_N}}\tilde{\mb{V}},\Pi^{\perp}_{(\Pi^{\perp}_{\mathrm{vec}(\mathbf{I}_N)}\tilde{\mb{V}})}\Pi^{\perp}_{\cvec{\mb{I}_N}}\mb{u}}$ share the same column space, the orthogonal projection matrix $\Pi^{\perp}_{\bs{\Delta}_s}$ can be expressed as:
\be
\label{sto_14}
\Pi^{\perp}_{\bs{\Delta}_s} = \Pi^{\perp}_{(\Pi^{\perp}_{\mathrm{vec}(\mathbf{I}_N)}\tilde{\mb{V}})} - \frac{1}{\mb{u}^H\Pi^{\perp}_{(\Pi^{\perp}_{\mathrm{vec}(\mathbf{I}_N)}\tilde{\mb{V}})}\mb{u}}\Pi^{\perp}_{(\Pi^{\perp}_{\mathrm{vec}(\mathbf{I}_N)}\tilde{\mb{V}})}\mb{u}\mb{u}^H\Pi^{\perp}_{(\Pi^{\perp}_{\mathrm{vec}(\mathbf{I}_N)}\tilde{\mb{V}})},
\ee 
that represents the \virg{semiparametric} analogous of \cite[eq. (14)]{Stoica_CRB}. 

To proceed, we need to evaluate the building blocks of the \eqref{sto_14}. Let us start with the orthogonal projection matrix: 
\be
\begin{split}
	\Pi^{\perp}_{(\Pi^{\perp}_{\mathrm{vec}(\mathbf{I}_N)}\tilde{\mb{V}})} &\triangleq \mb{I}_{N^2} - \Pi^{\perp}_{\mathrm{vec}(\mb{I}_N)}\mb{V}\mb{J}\left(\mb{J}^H\mb{V}^H \Pi^{\perp}_{ \mathrm{vec}(\mb{I}_N)} \mb{V}\mb{J} \right) ^{-1} \mb{V}^H\mb{J}^H \Pi^{\perp}_{\mathrm{vec}(\mathbf{I}_N)}\\
	&=\mb{I}_{N^2} - \Pi^{\perp}_{\mathrm{vec}(\mb{I}_N)}\mb{V}\left(\mb{V}^H \Pi^{\perp}_{ \mathrm{vec}(\mb{I}_N)} \mb{V} \right) ^{-1} \mb{V}^H \Pi^{\perp}_{\mathrm{vec}(\mathbf{I}_N)}\\
	&=\Pi^{\perp}_{(\Pi^{\perp}_{\mathrm{vec}(\mathbf{I}_N)}\mb{V})}.
\end{split}
\ee

The first term that we are going to evaluate is $\left(\mathbf{V}^H \Pi^{\perp}_{ \mathrm{vec}(\mathbf{I}_N)} \mathbf{V} \right) ^{-1}$. We have that:
\be
\begin{split}
	&\left(\mathbf{V}^H \Pi^{\perp}_{ \mathrm{vec}(\mathbf{I}_N)} \mathbf{V} \right) ^{-1} \\
	&=\quadre{(\bs{\Sigma}_0^{-T/2}\mb{A}^* \otimes \bs{\Sigma}_0^{-1/2}\mb{A})^H\Pi^{\perp}_{\mathrm{vec}(\mathbf{I}_N)}(\bs{\Sigma}_0^{-T/2}\mb{A}^* \otimes \bs{\Sigma}_0^{-1/2}\mb{A})}^{-1}\\
	&=\quadre{(\mb{A}^T\bs{\Sigma}_0^{-*/2} \otimes \mb{A}^H\bs{\Sigma}_0^{-H/2})\tonde{\mathbf{I}_{N^2} - \frac{1}{N}\mathrm{vec}(\mathbf{I}_N)\mathrm{vec}(\mathbf{I}_N)^T}(\bs{\Sigma}_0^{-T/2}\mb{A}^* \otimes \bs{\Sigma}_0^{-1/2}\mb{A})}^{-1}\\
	&= \left[ (\mb{A}^T\bs{\Sigma}_0^{-*/2} \otimes \mb{A}^H\bs{\Sigma}_0^{-H/2})(\bs{\Sigma}_0^{-T/2}\mb{A}^* \otimes \bs{\Sigma}_0^{-1/2}\mb{A})- \times \right. \\
	&\left. \times - N^{-1}((\bs{\Sigma}_0^{-H/2}\mb{A})^T \otimes \mb{A}^H\bs{\Sigma}_0^{-H/2})\mathrm{vec}(\mathbf{I}_N)\mathrm{vec}(\mathbf{I}_N)^T(\bs{\Sigma}_0^{-T/2}\mb{A}^* \otimes (\mb{A}^T\bs{\Sigma}_0^{-T/2})^T)\right]^{-1} \\
	&= \quadre{((\mb{A}^H\bs{\Sigma}_0^{-1}\mb{A})^* \otimes \mb{A}^H\bs{\Sigma}_0^{-1}\mb{A}) - N^{-1}\cvec{\mb{A}^H\bs{\Sigma}_0^{-1}\mb{A}}\cvec{\mb{A}^H\bs{\Sigma}_0^{-1}\mb{A}}^H}^{-1}\\
	& = ((\mb{A}^H\bs{\Sigma}_0^{-1}\mb{A})^* \otimes \mb{A}^H\bs{\Sigma}_0^{-1}\mb{A})^{-1} + N^{-1} \frac{\cvec{(\mb{A}^H\bs{\Sigma}_0^{-1}\mb{A})^{-1}}\cvec{(\mb{A}^H\bs{\Sigma}_0^{-1}\mb{A})^{-1}}^H}{1-N^{-1}\mathrm{tr}(\mb{I}_{K})}\\
	& = ((\mb{A}^H\bs{\Sigma}_0^{-1}\mb{A})^* \otimes \mb{A}^H\bs{\Sigma}_0^{-1}\mb{A})^{-1} + \frac{\cvec{(\mb{A}^H\bs{\Sigma}_0^{-1}\mb{A})^{-1}}\cvec{(\mb{A}^H\bs{\Sigma}_0^{-1}\mb{A})^{-1}}^H}{N-K}\\
	& = ((\mb{A}^H\bs{\Sigma}_0^{-1}\mb{A})^{-T/2} \otimes (\mb{A}^H\bs{\Sigma}_0^{-1}\mb{A})^{-1/2})\tonde{\mb{I}_{K^2} +\frac{1}{N-K}\cvec{\mb{I}_K}\cvec{\mb{I}_K}^T} \times \\
	& \times ((\mb{A}^H\bs{\Sigma}_0^{-1}\mb{A})^{-T/2} \otimes (\mb{A}^H\bs{\Sigma}_0^{-1}\mb{A})^{-1/2}),
\end{split}
\ee
where we use the fact that $\mb{A}^H\bs{\Sigma}_0^{-1}\mb{A}$ is Hermitian. Moreover, we have that:
\be
\begin{split}
	\Pi^{\perp}_{ \mathrm{vec}(\mathbf{I}_N)} \mathbf{V} &= \left(\mb{I}_{N^2} - \frac{1}{N}\mathrm{vec}(\mb{I}_N)\mathrm{vec}(\mb{I}_N)^T \right)(\bs{\Sigma}_0^{-T/2}\mb{A}^* \otimes \bs{\Sigma}_0^{-1/2}\mb{A})\\
	& = (\bs{\Sigma}_0^{-T/2}\mb{A}^* \otimes \bs{\Sigma}_0^{-1/2}\mb{A}) - \frac{1}{N}\mathrm{vec}(\mb{I}_N) \cvec{\mb{A}^H\bs{\Sigma}_0^{-1}\mb{A}}^H.
\end{split}
\ee

At this point, let us define the matrix $\mb{B}$ as:
\be
\mb{B} \triangleq [(\mb{X}^H\mb{X})^{-1/2}\mb{X}^H]^T \otimes \mb{X}(\mb{X}^H\mb{X})^{-1/2},
\ee
where 
\be
\mb{X} \triangleq \bs{\Sigma}_0^{-1/2}\mb{A}.
\ee

Finally, putting the previous results together, we have:
\be
\begin{split}
	\Pi^{\perp}_{\mathrm{vec}(\mathbf{I}_N)}&\mathbf{V}\left(\mathbf{V}^H \Pi^{\perp}_{ \mathrm{vec}(\mathbf{I}_N)} \mathbf{V} \right) ^{-1} \mathbf{V}^H \Pi^{\perp}_{\mathrm{vec}(\mathbf{I}_N)}\\
	& = \tonde{\mb{B}-\frac{1}{N}\cvec{\mb{I}_N}\cvec{\mb{I}_K}^T}\tonde{\mb{I}_{K^2} +\frac{1}{N-K}\cvec{\mb{I}_K}\cvec{\mb{I}_K}^T} \times \\
	&\times \tonde{\mb{B}-\frac{1}{N}\cvec{\mb{I}_N}\cvec{\mb{I}_K}^T}^H\\
	& = \tonde{\mb{B} +\frac{1}{N-K}\cvec{\Pi_{\mb{X}}}\cvec{\mb{I}_K}^T -\frac{1}{N-K}\cvec{\mb{I}_N}\cvec{\mb{I}_K}^T} \times \\
	&\times \tonde{\mb{B}^H-\frac{1}{N}\cvec{\mb{I}_K}\cvec{\mb{I}_N}^T}\\
	& = \mb{B}\mb{B}^H+\frac{1}{N-K}\cvec{\Pi_{\mb{X}}}\cvec{\Pi_{\mb{X}}}^H - \frac{1}{N-K}\cvec{\mb{I}_N}\cvec{\Pi_{\mb{X}}}^H \times\\
	& \times -\frac{1}{N}\cvec{\Pi_{\mb{X}}}\cvec{\mb{I}_N}^T -\frac{K}{N(N-K)}\cvec{\Pi_{\mb{X}}}\cvec{\mb{I}_N}^T \times\\
	& \times  +\frac{K}{N(N-K)}\cvec{\mb{I}_N}\cvec{\mb{I}_N}^T\\
	& = \Pi_{\mb{X}}^T \otimes \Pi_{\mb{X}}+\frac{1}{N-K}\cvec{\Pi_{\mb{X}}}\cvec{\Pi_{\mb{X}}}^H - \frac{1}{N-K}\cvec{\mb{I}_N}\cvec{\Pi_{\mb{X}}}^H \times\\
	& \times -\frac{1}{N-K}\cvec{\Pi_{\mb{X}}}\cvec{\mb{I}_N}^T  +\frac{K}{N(N-K)}\cvec{\mb{I}_N}\cvec{\mb{I}_N}^T.
\end{split}
\ee 

Moreover, we have that:
\be
\begin{split}
	\Pi^{\perp}_{(\Pi^{\perp}_{\mathrm{vec}(\mathbf{I}_N)}\mathbf{V})} \triangleq & \mathbf{I}_{N^2} -\Pi_{\mb{X}}^T \otimes \Pi_{\mb{X}}-\frac{1}{N-K}\cvec{\Pi_{\mb{X}}}\cvec{\Pi_{\mb{X}}}^H +\\
	&+ \frac{1}{N-K}\cvec{\mb{I}_N}\cvec{\Pi_{\mb{X}}}^H +\\
	& +  \frac{1}{N-K}\cvec{\Pi_{\mb{X}}}\cvec{\mb{I}_N}^T - \frac{K}{N(N-K)}\cvec{\mb{I}_N}\cvec{\mb{I}_N}^T.
\end{split}
\ee

The basic tools to evaluate the SSCRB in (62) are now ready to be used. Let's start by evaluate the matrix product $\Pi^{\perp}_{\bs{\Delta}_s}\mb{G}_s$. To this end, let us consider the \virg{column-wise} version of $\mb{G}_s$, i.e.:
\be
\label{mat_per_G}
\Pi^{\perp}_{\bs{\Delta}_s}\mb{G}_s = \quadre{\Pi^{\perp}_{\bs{\Delta}_s}\Pi^{\perp}_{\mathrm{vec}(\mathbf{I}_N)}\mb{g}_1,\cdots,\Pi^{\perp}_{\bs{\Delta}_s}\Pi^{\perp}_{\mathrm{vec}(\mathbf{I}_N)}\mb{g}_K},
\ee
and the $k$-th vector $\mb{g}_k$ is the one defined in \cite[eq. (17)]{Stoica_CRB} as:
\be
\label{sto_17}
\mb{g}_k \triangleq \cvec{\mb{Z}_k + \mb{Z}_k^H},
\ee
where
\be
\label{sto_18}
\mb{Z}_k \triangleq \mb{X}\mb{c}_{0,k}\mb{d}_{0,k}^H\bs{\Sigma}_0^{-1/2},
\ee
and 
\be
\label{der_steer}
\mb{d}_{0,k} \triangleq \left. \frac{d\mb{a}(\nu_k)}{d\nu_k} \right|_{\nu_k = \nu_{0,k}},
\ee
while $\mb{c}_{0,k}$ represents the $k$-th column of the signal covariance matrix $\bs{\Gamma}_0$, such that $\bs{\Gamma}_0 = [\mb{c}_{0,1},\cdots,\mb{c}_{0,K}]$. Given \eqref{sto_14}, in order to evaluate the $k$-th column of the matrix \eqref{mat_per_G}, we have to calculate the following matrix:
\be
\label{sto_14_plus}
\begin{split}
	\Pi^{\perp}_{\bs{\Delta}_s}\Pi^{\perp}_{\mathrm{vec}(\mathbf{I}_N)}\mb{g}_k & = \Pi^{\perp}_{(\Pi^{\perp}_{\mathrm{vec}(\mathbf{I}_N)}\mb{V})}\Pi^{\perp}_{\mathrm{vec}(\mathbf{I}_N)}\mb{g}_k -\\
	&- \frac{1}{\mb{u}^H\Pi^{\perp}_{(\Pi^{\perp}_{\mathrm{vec}(\mathbf{I}_N)}\mb{V})}\mb{u}}\Pi^{\perp}_{(\Pi^{\perp}_{\mathrm{vec}(\mathbf{I}_N)}\mb{V})}\mb{u}\mb{u}^H\Pi^{\perp}_{(\Pi^{\perp}_{\mathrm{vec}(\mathbf{I}_N)}\mb{V})} \Pi^{\perp}_{\mathrm{vec}(\mathbf{I}_N)}\mb{g}_k.
\end{split}
\ee 

The first term in \eqref{sto_14_plus} can be evaluated as:
\be
\begin{split}
	\Pi^{\perp}_{(\Pi^{\perp}_{\mathrm{vec}(\mathbf{I}_N)}\mathbf{V})}&\Pi^{\perp}_{\mathrm{vec}(\mathbf{I}_N)}\mb{g}_k = 	\Pi^{\perp}_{\mathrm{vec}(\mathbf{I}_N)}\mb{g}_k - \Pi^{\perp}_{\mathrm{vec}(\mathbf{I}_N)}\mathbf{V}\left(\mathbf{V}^H \Pi^{\perp}_{ \mathrm{vec}(\mathbf{I}_N)} \mathbf{V} \right) ^{-1} \mathbf{V}^H \Pi^{\perp}_{\mathrm{vec}(\mathbf{I}_N)}\mb{g}_k\\
	&=\Pi^{\perp}_{\mathrm{vec}(\mathbf{I}_N)}\Pi^{\perp}_{(\Pi^{\perp}_{\mathrm{vec}(\mathbf{I}_N)}\mathbf{V})}\mb{g}_k\\
	& =\Pi^{\perp}_{\mathrm{vec}(\mathbf{I}_N)}\tonde{\mb{g}_k - \cvec{\Pi_{\mb{X}}(\mb{Z}_k + \mb{Z}_k^H)\Pi_{\mb{X}}}}\\
	& = \Pi^{\perp}_{\mathrm{vec}(\mathbf{I}_N)}\tonde{\cvec{\mb{Z}_k + \mb{Z}_k^H} - \cvec{\Pi_{\mb{X}}\mb{Z}_k\Pi_{\mb{X}} + \Pi_{\mb{X}}{\mb{Z}}_k^H\Pi_{\mb{X}}}}\\
	& = \Pi^{\perp}_{\mathrm{vec}(\mathbf{I}_N)}\tonde{\cvec{\mb{Z}_k + \mb{Z}_k^H - \mb{Z}_k\Pi_{\mb{X}} - \Pi_{\mb{X}}{\mb{Z}}_k^H}}\\
	& = \Pi^{\perp}_{\mathrm{vec}(\mathbf{I}_N)}\cvec{\mb{Z}_k\Pi_{\mb{X}}^{\perp} + \Pi_{\mb{X}}^{\perp}\mb{Z}_k^H},
\end{split}
\ee
where, according to the properties of the projection matrices, $\Pi_{\mb{X}} = \Pi_{\mb{X}}^H$ and, by definition of the matrix $\mb{Z}_k$, we have that $\Pi_{\mb{X}}\mb{Z}_k=\mb{Z}_k$. Moreover, the last equality follows from the fact that:
\be
\begin{split}
	\Pi^{\perp}_{\mathrm{vec}(\mathbf{I}_N)} &\cvec{\mb{Z}_k\Pi_{\mb{X}}^{\perp} + \Pi_{\mb{X}}^{\perp}\mb{Z}_k^H}\\
	& = \cvec{\mb{Z}_k\Pi_{\mb{X}}^{\perp} + \Pi_{\mb{X}}^{\perp}\mb{Z}_k^H} - N^{-1}\cvec{\mb{I}_N}\cvec{\mb{I}_N}^T\cvec{\mb{Z}_k\Pi_{\mb{X}}^{\perp} + \Pi_{\mb{X}}^{\perp}\mb{Z}_k^H}\\
	& = \cvec{\mb{Z}_k\Pi_{\mb{X}}^{\perp} + \Pi_{\mb{X}}^{\perp}\mb{Z}_k^H} - N^{-1}\mathrm{tr}(\mb{Z}_k\Pi_{\mb{X}}^{\perp} + \Pi_{\mb{X}}^{\perp}\mb{Z}_k^H)\cvec{\mb{I}_N}\\
	& = \cvec{\mb{Z}_k\Pi_{\mb{X}}^{\perp} + \Pi_{\mb{X}}^{\perp}\mb{Z}_k^H} - N^{-1}\quadre{\mathrm{tr}(\Pi_{\mb{X}}^{\perp}\mb{Z}_k) + \mathrm{tr}(\Pi_{\mb{X}}^{\perp}\mb{Z}_k)^H}\cvec{\mb{I}_N}\\
	& = \cvec{\mb{Z}_k\Pi_{\mb{X}}^{\perp} + \Pi_{\mb{X}}^{\perp}\mb{Z}_k^H},	
\end{split}
\ee
where we used the fact that, by definition of the matrices $\mb{Z}_k$ and $\Pi_{\mb{X}}^{\perp}$, we have that $\Pi_{\mb{X}}^{\perp}\mb{Z}_k = \mb{0}$.

The second term in subtraction in \eqref{sto_14_plus} is nil since
\be
\begin{split}
	\mb{u}^H\Pi^{\perp}_{(\Pi^{\perp}_{\mathrm{vec}(\mathbf{I}_N)}\mathbf{V})}&\Pi^{\perp}_{\mathrm{vec}(\mathbf{I}_N)}\mb{g}_k  = \cvec{\bs{\Sigma}_0^{-1}}^H\cvec{\mb{Z}_k\Pi_{\mb{X}}^{\perp} + \Pi_{\mb{X}}^{\perp}\mb{Z}_k^H}\\
	& = \mathrm{tr}(\bs{\Sigma}_0^{-1}(\mb{Z}_k\Pi_{\mb{X}}^{\perp} + \Pi_{\mb{X}}^{\perp}\mb{Z}_k^H)) = 2\mathrm{Re}\{\mathrm{tr}(\Pi_{\mb{X}}^{\perp}\bs{\Sigma}_0^{-1}\mb{Z}_k)\} = 0,
\end{split}
\ee
where we used the fact that the matrix $\mb{X}$ and $\bs{\Sigma}_0^{-1}\mb{Z}_k$ share the same column space (see also \cite[eqs. (25), (26) and (27)]{Stoica_CRB}). We're almost done now. The only thing that is left to be done is to substitute the obtained results in (62). Specifically:
\be
\label{sto_30}
\begin{split}
	\frac{N(N+1)}{L\bar{E}\{\mathcal{Q}^2\psi_0(\mathcal{Q})^2\}}&\quadre{\mathrm{SCRB}(\bs{\nu}_0|\bs{\zeta}_0,\sigma_0^2, h_0)}^{-1}_{i,j} = \mb{g}_i^H\Pi^{\perp}_{\mathrm{vec}(\mathbf{I}_N)} \Pi^{\perp}_{(\Pi^{\perp}_{\mathrm{vec}(\mathbf{I}_N)}\mathbf{V})}\Pi^{\perp}_{\mathrm{vec}(\mathbf{I}_N)}\mb{g}_j \\
	&= \mb{g}_i^H  \Pi^{\perp}_{(\Pi^{\perp}_{\mathrm{vec}(\mathbf{I}_N)}\mathbf{V})}\Pi^{\perp}_{\mathrm{vec}(\mathbf{I}_N)}\mb{g}_j\\
	&=\mb{g}_i^H\cvec{\mb{Z}_j\Pi_{\mb{X}}^{\perp} + \Pi_{\mb{X}}^{\perp}\mb{Z}_j^H}\\
	& = \mathrm{tr}(\mb{Z}_i\mb{Z}_j\Pi_{\mb{X}}^{\perp} + \mb{Z}_i\Pi_{\mb{X}}^{\perp}\mb{Z}_j^H + \mb{Z}_i^H\mb{Z}_j\Pi_{\mb{X}}^{\perp}+\mb{Z}_i^H\Pi_{\mb{X}}^{\perp}\mb{Z}_j^H)\\
	&=2\mathrm{Re}\graffe{\mathrm{tr}(\mb{Z}_i\Pi_{\mb{X}}^{\perp}\mb{Z}_j^H)}, \quad i,j = 1,\ldots,K,
\end{split}
\ee
that represents exactly \cite[eq. 30]{Stoica_CRB}, except for a scalar term. As before, we used the equality $\Pi_{\mb{X}}^{\perp}\mb{Z}_{i} = \mb{0}$.

To conclude, by substituting in \eqref{sto_30} the expression of the matrix $\mb{Z}_k$ given in \eqref{sto_18} and by using \cite[eq. (30)]{Stoica_CRB}, we get the expression of the SSCRB reported in (62).


\bibliographystyle{IEEEtran}
\bibliography{ref_unico_complessi}

\end{document}